\documentclass[aps,prx,superscriptaddress,amsmath,twocolumn,amssymb]{revtex4-2}

\usepackage{subfigure}
\usepackage{amsmath}
\usepackage{pgfplots}
\pgfplotsset{compat=1.5}
\usepgfplotslibrary{groupplots}

\usepackage{color}

\usepackage{xcolor}
\usepackage{bm}
\usepackage{multirow}
\usepackage{placeins}
\usepackage{soul}
\usepackage{color,xcolor}
\usepackage[colorlinks,linkcolor=red,anchorcolor=blue,citecolor=blue,urlcolor=blue]{hyperref}
\usepackage{cleveref}
\usepackage{amsmath}

\newcommand{\rg}{r_{\mathrm{g}}}
\newcommand{\rgp}{r_{\mathrm{g}}'}
\newcommand{\nrg}{n_{\mathrm{g}}}

\begin{document}

\title{Truncated hybrid tensor networks for distributed quantum simulation}

\author{Yong Liu}
\email{liuyong09@nudt.edu.cn}
\affiliation{College of Computer Science and Technology, National University of Defense Technology, Changsha 410073, China}

\author{Guangyao Huang}
\affiliation{College of Computer Science and Technology, National University of Defense Technology, Changsha 410073, China}

\author{Weixu Shi}
\affiliation{College of Computer Science and Technology, National University of Defense Technology, Changsha 410073, China}

\author{Yizhi Wang}
\affiliation{College of Computer Science and Technology, National University of Defense Technology, Changsha 410073, China}

\author{Jiandong Ouyang}
\affiliation{College of Computer Science and Technology, National University of Defense Technology, Changsha 410073, China}

\author{Zeqian Chen}
\affiliation{College of Computer Science and Technology, National University of Defense Technology, Changsha 410073, China}

\author{Junjie Wu}
\email{junjiewu@nudt.edu.cn}
\affiliation{College of Computer Science and Technology, National University of Defense Technology, Changsha 410073, China}

\begin{abstract}
Simulation of quantum many-body systems is a principal application of quantum computing, but available devices remain limited by the number of qubits and cannot accommodate systems of the desired size.
One approach is to host a large evolution across several modest distributed systems that exchange classical information alone.
Current protocols such as circuit knitting incur a quasi-probability sampling cost that grows exponentially with the number of remote gates across the cut, even when physical correlations across the cut are actually bounded.
In this work, we present a truncated hybrid tensor network (THTN) framework in which a remote two-body gate is applied as a sum of local unitaries on the subsystems, linked by a classical connecting tensor, and an interface truncation retains Schmidt modes across the cut.
Remote gates between the same subsystems can be merged on one connector, so non-one-dimensional models, such as layered partitions with strong intra-layer and weaker interlayer couplings, may be cast onto a one-dimensional cut.
The retained non-negative Schmidt coefficients also define a sampling rule for local observables.
We validate this distributed protocol by classical simulation against time-evolving block decimation (TEBD) at the same bond dimension.
The truncated dynamics track the TEBD references on chains and ladders, with the clearest gain on a layered model.
\end{abstract}

\date{\today}

\maketitle

\section{Introduction}

Simulating the dynamics of quantum many-body systems is among the principal motives for quantum computation~\cite{Feynman1982}.
Beyond condensed-matter models of correlated phases and nonequilibrium dynamics~\cite{Anderson1984,Sachdev2023,Monroe2021}, the same task arises in quantum chemistry~\cite{Cao2019,McArdle2020}, in lattice-gauge and high-energy settings~\cite{Jordan2012,Zohar2016}, and as a structured benchmark for early fault-tolerant processors~\cite{Gao2025,Acharya2025}.

As the number of lattice sites grows, more qubits are required and more entanglement is generated under unitary time evolution.
Present devices remain limited by the number of qubits for a lattice of interest.
The evolution can then be hosted across several modest processors linked by classical communication only~\cite{Monroe2013,Wehner2018}.
The lattice must therefore be partitioned into subsystems.
Unitary time evolution is commonly approximated by Trotter product formulas~\cite{Suzuki1991,Childs2021}, so each step combines local updates with remote two-body Pauli gates across subsystem boundaries.
Circuit knitting and related cutting protocols~\cite{Peng2020,Mitarai2021,Piveteau2024,Gentinetta2024,Harrow2025} implement those gates by channel-level decompositions and quasi-probability sampling.
The associated sampling overhead grows exponentially with the number of remote gates across the cut.
The issue is especially relevant when physical correlations across the cut are actually bounded, as in many one- and quasi-one-dimensional models~\cite{Hastings2007,Eisert2010}.
Layered geometries with strong intra-layer and weaker interlayer couplings furnish a sharp instance of that scale separation, as in van der Waals heterostructures (e.g., bilayer graphene)~\cite{Geim2013,McCann2013,Kennes2021,Cao2018}.
Other methods such as deep variational quantum eigensolver (deep VQE)~\cite{Fujii2022} and entanglement forging~\cite{Eddins2022} address a similar partitioned-hardware setting for ground states.

Hybrid tensor networks (HTNs)~\cite{Yuan2021,Mello2024,Kanno2021} offer one framework for this distributed setting.
They encode a global state on sites that may hold classical coefficients or local quantum states, and can run conventional tensor-network updates with quantum devices~\cite{Liu2025}.
Existing constructions are largely static, aimed at state representation or variational preparation rather than at evolution under successive remote Trotter gates.
Extending HTNs to partitioned digital dynamics therefore requires both a dynamical update for cross-boundary two-body gates and a truncation mechanism compatible with hybrid sites.
Neither a matrix-product-state (MPS) bond singular-value decomposition nor quantum SVD (QSVD)~\cite{Rebentrost2018,Wang2021,Carlos2020} applies directly, because each site mixes classical coefficients with local physical states.

\begin{figure*}
  \centering
  \includegraphics[width=\textwidth]{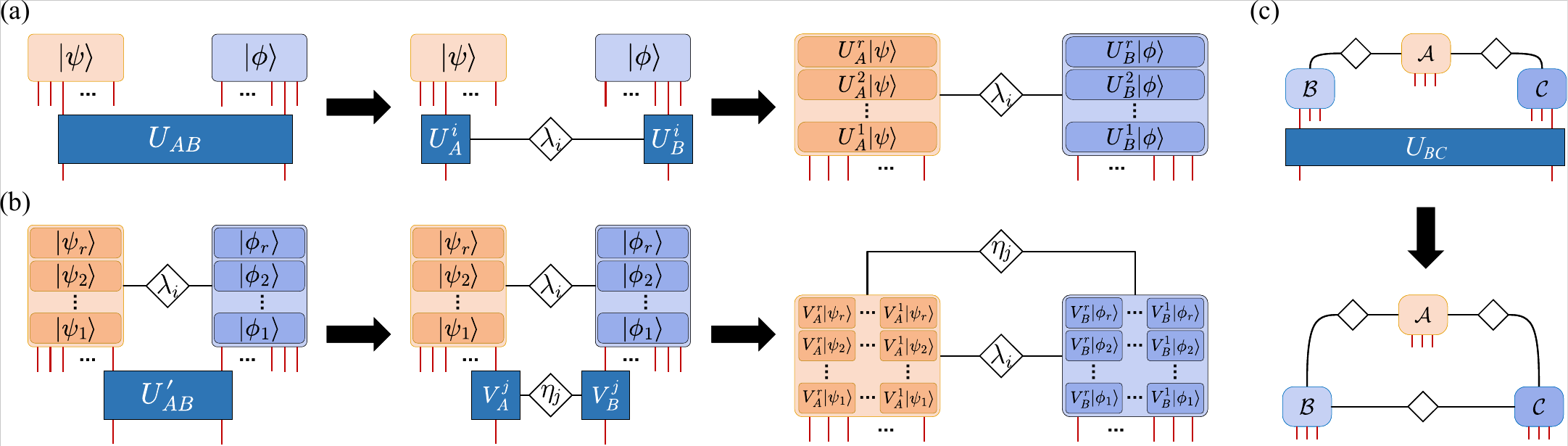}
  \caption{Nonlocal two-qubit gates on a distributed system in the THTN representation. (a) A first remote gate introduces the connecting tensor $\boldsymbol{\lambda}=\mathrm{Diag}(\lambda_1,\ldots,\lambda_{\rg})$ on a new bond, and each hybrid tensor gains a classical index. (b) A further gate on an already connected pair appends $\boldsymbol{\eta}=\mathrm{Diag}(\eta_1,\ldots,\eta_{\rgp})$. Flattening the enlarged site indices gives $\boldsymbol{\lambda}\otimes\boldsymbol{\eta}$. Within each subsystem the THTN representation does not track which physical qubit the local unitary acts on. (c) An example of more general THTN connector topology. Three hybrid sites $[\mathcal{A}]$, $[\mathcal{B}]$, and $[\mathcal{C}]$ are pairwise linked by classical connectors, showing that remote gates can place bonds beyond chain or tree layouts.
  }\label{fig_NonLocalGate}
\end{figure*}

For distributed quantum simulation with time evolution, we introduce a truncated hybrid tensor network (THTN).
A remote two-body gate is applied as a sum of local unitaries on the hybrid sites, with classical coefficients stored as a connecting tensor.
Cross-boundary couplings thereby enter the network explicitly as evolution proceeds.
An interface truncation then retains the leading Schmidt modes on each connector, in the spirit of bond truncation in conventional tensor networks.
It keeps the bond dimension under control when correlations across the cut remain bounded, without compressing entanglement within each subsystem.
Remote gates between the same subsystems can be merged on one connector, so non-one-dimensional models may be cast onto a one-dimensional cut.
Observables are obtained by sampling connector indices according to the resulting non-negative Schmidt coefficients and accumulating the corresponding shot contributions.
Numerical benchmarks against time-evolving block decimation (TEBD)~\cite{Vidal2004,Schollwock2011} at the same bond dimension $\chi$ indicate that truncation at the cut maintains high simulation accuracy relative to TEBD.
The advantage is clearest on a layered partition that evolves strong intra-layer correlations within the hybrid sites and truncates only the weaker interlayer connector.

This paper is organized as follows. Section~\ref{sec:framework} develops the THTN framework. It defines hybrid sites and classical connectors, the remote two-body gate protocol, the method for truncating connecting interfaces between hybrid tensor sites, and gives the sampling rule for observables. It then closes with the passage from distributed Trotter simulation to distributed quantum computation. Section~\ref{sec:numerical} reports numerical benchmarks against TEBD on one-dimensional, quasi-one-dimensional, and layered systems. Section~\ref{sec:discussion} discusses limitations and outlook.

\section{Framework for distributed quantum simulation}\label{sec:framework}

\subsection{Hybrid tensors}

A hybrid tensor site is a collection of data units, each of which may be a classical scalar or a quantum state.
It takes the form
\begin{equation}\label{eq:htn}
  [\mathcal{A}]_{\mathbf{c}}^{\mathbf{q}} = w_{c_1c_2\ldots}|\psi_{c_1c_2\ldots}\rangle
\end{equation}
where the subscripts $\mathbf{c}=\{c_1,c_2,\ldots\}$ are classical indices that specify the site shape and the superscripts $\mathbf{q}=\{q_1,q_2,\ldots\}$ are quantum indices labeling the constituent qubits.
Each entry is an $n$-qubit state $|\psi_{c_1c_2\ldots}\rangle$ multiplied by a weight $w_{c_1c_2\ldots}$, initialized to 1.
These weights are used to track the norm change during the hybrid contraction.
As in a classical tensor network, hybrid tensors are linked through bond indices.
Contracting all hybrid tensors over those indices yields a global state.

Contractions of hybrid tensors follow the same logic as classical tensor contractions, with additional rules for index type.
We summarize the contractions in ten cases, depending on the tensors and index types involved (see the Supplemental Material for the full list, circuit implementation, and state evolution).
As an example, contracting a classical tensor $C_{ij}$ with a hybrid tensor $w_k|\psi_k\rangle$ on its classical index is summarized as a Case~2 contraction and gives $|\phi_j\rangle=\sum_{i}w_i C_{ij}|\psi_i\rangle$.
The resulting state should in general be renormalized, introducing a nontrivial weight.

\subsection{Remote two-body gates}

In Trotter simulation, the remote operations required across a partition are two-body terms $e^{-iH\Delta t}$ with $H$ a combination of Pauli strings, generally $H=\lambda_0 XX+\lambda_1 YY+\lambda_2 ZZ$.
When $H$ couples qubits on different subsystems, the term is implemented as a remote gate.

As for any entangling operation on distributed subsystems, such a gate requires additional classical processing.
A two-qubit unitary can be decomposed into a sum of local operations:
\begin{equation}\label{eq:decomp}
U=\sum_{i=1}^{\rg} \lambda_i [U_A]_i\otimes [U_B]_i,
\end{equation}
where $[U_A]_i$ and $[U_B]_i$ are local unitaries and $\lambda_i$ are classical coefficients~\cite{Bravyi2016}.

The THTN implementation of such a remote gate is shown in Fig.~\ref{fig_NonLocalGate}. Applying a remote gate between two subsystems gives
\begin{equation}\label{eq:apply}
  U|\psi\rangle_A|\phi\rangle_B = \sum_i \lambda_i \,[U_A]_i|\psi\rangle_A \otimes [U_B]_i|\phi\rangle_B,
\end{equation}
producing two hybrid tensors
\begin{equation}\label{eq:three_hybrid_tensor}
\begin{aligned}
  \relax[\mathcal{A}]_i &= [U_A]_i|\psi\rangle_A,\\
  \relax[\mathcal{B}]_i &= [U_B]_i|\phi\rangle_B,
\end{aligned}
\end{equation}
and a classical {\em connecting tensor} (connector)~$\boldsymbol{\lambda}$, as shown in Fig.~\ref{fig_NonLocalGate}(a).
We write $\boldsymbol{\lambda}=\mathrm{Diag}(\lambda_1,\ldots,\lambda_{\rg})$.
The gate-decomposition rank $\rg$ is the bond dimension introduced.
If the subsystems are already connected, the gate enlarges the hybrid-site indices and appends a second connector
\begin{equation}\label{eq:three_hybrid_tensor_2}
\begin{aligned}
  \relax[\mathcal{A}']_{ij} &= [U_A]_j[\mathcal{A}]_{i},\\
  \relax[\mathcal{B}']_{ij} &= [U_B]_j[\mathcal{B}]_{i},
\end{aligned}
\end{equation}
with $\boldsymbol{\eta}=\mathrm{Diag}(\eta_1,\ldots,\eta_{\rgp})$, as shown in Fig.~\ref{fig_NonLocalGate}(b).
Flattening the enlarged sites and merging the connectors gives a combined bond with tensor $\boldsymbol{\lambda}\otimes \boldsymbol{\eta}$.
The THTN layout is therefore not tied to a chain or tree.
Figure~\ref{fig_NonLocalGate}(c) shows a three-subsystem example in which every pair is connected by a classical connector.
In practice, however, such non-tree layouts can sometimes be reduced to a one-dimensional cut.
This is possible because, within each subsystem, the THTN representation does not track which physical qubit a local unitary acts on.
It only records which subsystems are coupled.
Remote gates that act between the same pair of subsystems can therefore be merged on a single connector, which may remove loops that would otherwise appear in the graph.

This update rule is the THTN analog of a simple update in tensor-network algorithms~\cite{Jiang2008}.
The bond-dimension increment is set by the gate-decomposition rank $\rg$ in Eq.~\eqref{eq:decomp}.
For Pauli-channel Trotter gates, $\rg$ equals the number of independent Pauli terms in $H$ plus 1 (including the identity channel).
Applying a remote gate for an Ising $XX$ term therefore enlarges the bond dimension on that cut by a factor of $2$.
A remote gate for a full XXZ term enlarges it by a factor of $4$.

\subsection{Truncating the bond dimensions}\label{sec:truncation}

Remote gates increase the bond dimension of THTNs, so controlled approximation requires truncating correlations at subsystem interfaces.
In classical tensor networks, that step is a singular-value decomposition of the local state, discarding components tied to small singular values.
Hybrid sites mix local physical states with classical coefficients, so no such local decomposition exists.
We therefore develop a compression method for HTNs.
The method has two steps, orthogonalization and connector decomposition, as shown in Fig.~\ref{fig:orth}.
When correlations across the cut remain bounded, the retained bond dimension can stay moderate while accuracy is preserved.

\subsubsection{Orthogonalization}

\begin{figure}
  \centering
  \includegraphics[width=0.4\textwidth]{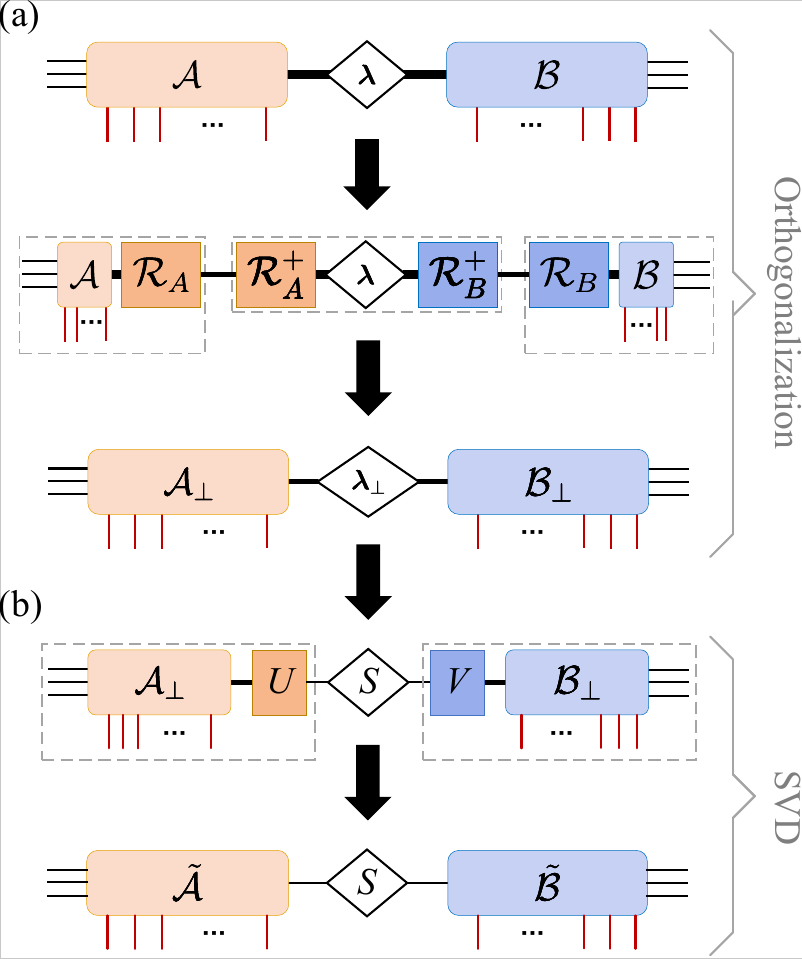}
  \caption{Orthogonalization and connector truncation at a partition cut. (a) Hybrid sites are orthonormalized along the cut index through the Gram matrix and mapping tensor $\mathcal{R}$. (b) The classical connecting tensor $\boldsymbol{\lambda}_\perp$ undergoes singular value decomposition. Its nonzero singular values equal the Schmidt coefficients of the bipartite state across the cut. Truncation keeps the largest $\chi$ singular values and the matching blocks of $U$ and $V$. After contracting them into the sites, the cut index of $\widetilde{\mathcal{A}}$ (and likewise of $\widetilde{\mathcal{B}}$) is reduced from dimension $d$ to $\chi$.}\label{fig:orth}
\end{figure}

Given a hybrid site $[\mathcal{A}]_{\mathbf{c}}^{\mathbf{q}}$ with classical indices $\mathbf{c}=\{c_1,c_2,\ldots\}$ and quantum indices $\mathbf{q}=\{q_1,q_2,\ldots\}$, we orthogonalize the constituent states along a classical index $i$ by Gram orthogonalization~\cite{Schuhmacher2024}, as illustrated in Fig.~\ref{fig:orth}(a).
This step introduces a classical {\em mapping tensor} on that index, together with an {\em unmapping tensor} that recovers the original basis and keeps the contracted HTN consistent.

We first form the Gram matrix with respect to the classical index $i$.
\begin{equation}\label{eq:gram_matrix}
\begin{aligned}
  G_{ii'} &= \sum_{\mathbf{c}^{-},\mathbf{q}}
                  [\mathcal{A}^*]^{\mathbf{q}}_{\mathbf{c}^{-},i}
                  [\mathcal{A}]^{\mathbf{q}}_{\mathbf{c}^{-},i'}\\
  &= \sum_{\mathbf{c}^{-},\mathbf{q}} \langle \psi_{\mathbf{c}^{-},i}|\mathbf{q}\rangle\langle \mathbf{q}|\psi_{\mathbf{c}^{-},i'}\rangle \\
  &= \sum_{\mathbf{c}^{-}} \langle \psi_{\mathbf{c}^{-},i}|\psi_{\mathbf{c}^{-},i'}\rangle
\end{aligned}
\end{equation}
where $\mathbf{c}^{-}$ denotes the classical indices other than $i$, and $d$ is the dimension of index $i$ (so $G\in\mathbb{C}^{d\times d}$).
Since $G$ is Hermitian, let $r=\mathrm{rank}(G)$ and write the spectral decomposition $G = P_r\Lambda_r P_r^{\dagger}$, where $\Lambda_r\in\mathbb{R}^{r\times r}$ contains the nonzero eigenvalues and $P_r\in\mathbb{C}^{d\times r}$ holds the corresponding orthonormal eigenvectors.
Theorem~1 shows how to orthogonalize the site while preserving the contracted network value.

{\bf Theorem 1}.
Given a hybrid site $[\mathcal{A}]_{\mathbf{c}}^{\mathbf{q}}=|\psi_\mathbf{c}\rangle$ with Gram matrix $G$, orthogonalization with respect to the index $i$ is achieved by
\begin{equation}\label{eq:orth}
  \relax[\mathcal{A}_\perp]^{\mathbf{q}}_{\mathbf{c}^{-},j} = \sum_i [\mathcal{A}]^{\mathbf{q}}_{\mathbf{c}^{-},i} \mathcal{R}_{ij},
\end{equation}
where the mapping tensor is $\mathcal{R}=P_r\Lambda_r^{-1/2}$.
It can be recovered by
\begin{equation}\label{eq:recover}
    \relax[\mathcal{A}]^{\mathbf{q}}_{\mathbf{c}^{-},j} = \sum_i [\mathcal{A}_\perp]^{\mathbf{q}}_{\mathbf{c}^{-},i} \mathcal{R}^+_{ij},
\end{equation}
where $\mathcal{R}^+=\Lambda_r^{1/2}P_r^\dagger$ is the unmapping tensor.

{\em Proof}.
The orthogonality condition
\begin{equation}
  \sum_{\mathbf{c}^{-},\mathbf{q}} [\mathcal{A}_\perp^*]^{\mathbf{q}}_{\mathbf{c}^{-},j}[\mathcal{A}_\perp]^{\mathbf{q}}_{\mathbf{c}^{-},k}=\delta_{jk},
\end{equation}
follows directly from the construction~\cite{Schuhmacher2024}.
The nontrivial claim is that recovery remains valid after truncating $\mathcal{R}$ to the $r$ modes with nonzero eigenvalues of $G$.
In that case $\mathcal{R}\mathcal{R}^+\neq I$ in general, and $\mathcal{R}$ has no proper right inverse.

In the truncation workflow, the weights of each hybrid-site branch are kept at $w=1$.
Collect the amplitudes of $[\mathcal{A}]_{\mathbf{c}}^{\mathbf{q}}$ into a matrix $A$ with entries
\begin{equation}\label{eq:conversion}
  A_{(\mathbf{c}^{-},\mathbf{q}),i}=[\mathcal{A}]^{\mathbf{q}}_{\mathbf{c}^{-},i}=\langle\mathbf{q}|\psi_{\mathbf{c}^{-},i}\rangle,
\end{equation}
so that rows are labeled by the composite index $(\mathbf{c}^{-},\mathbf{q})$ and columns by the cut index $i$.
Then Eq.~\eqref{eq:gram_matrix} reads $G=A^\dagger A$, with $\mathrm{rank}(A)=r$.

Write $M=P_rP_r^\dagger$, and a short calculation gives $\mathcal{R}\mathcal{R}^+=M$.
In matrix form, Eqs.~\eqref{eq:orth} and~\eqref{eq:recover} become $A_\perp=A\mathcal{R}$ and $A=A_\perp\mathcal{R}^+$.
Recovery therefore requires $A\mathcal{R}\mathcal{R}^+=A$, i.e., $AM=A$.
Because $M$ is Hermitian, it is enough to prove the equivalent identity $MA^\dagger=A^\dagger$.

A direct calculation gives
\begin{equation}
  MG =P_r P_r^\dagger \cdot P_r\Lambda_r P_r^\dagger
  =P_r\Lambda_r P_r^\dagger
  =G,
\end{equation}
where we used $P_r^\dagger P_r=I_r$.
Let $\mathbf{col}(\cdot)$ denote the column space of a matrix.
Then $M$ acts as the identity on $\mathbf{col}(G)$: for every $\mathbf{v}\in\mathbf{col}(G)$ there exists $\mathbf{x}$ with $G\mathbf{x}=\mathbf{v}$, and
\begin{equation}
  M\mathbf{v}=MG\mathbf{x}=G\mathbf{x}=\mathbf{v}.
\end{equation}
For every $\mathbf{x}\in\mathbb{C}^d$ one also has
\begin{equation}
G\mathbf{x}=A^\dagger A\mathbf{x}\in\mathbf{col}(A^\dagger),
\end{equation}
so $\mathbf{col}(G)\subseteq\mathbf{col}(A^\dagger)$.
The ranks match, $\mathrm{rank}(G)=\mathrm{rank}(A)=r=\mathrm{rank}(P_r)=\mathrm{rank}(M)$, and therefore
\begin{equation}
\mathbf{col}(G)=\mathbf{col}(A^\dagger).
\end{equation}
Consequently every column of $A^\dagger$ lies in $\mathbf{col}(G)$, on which $M$ acts as the identity. Hence
\begin{equation}
  MA^\dagger=A^\dagger.
\end{equation}
Therefore, we have $AM=A\mathcal{R}\mathcal{R}^+=A$.
The truncated recovery map restores the original hybrid tensor. $\blacksquare$

\subsubsection{Truncation on the connecting dimension}

Given two hybrid tensors $[\mathcal{A}]_{\mathbf{c}_A}^{\mathbf{q}_A}$ and $[\mathcal{B}]_{\mathbf{c}_B}^{\mathbf{q}_B}$ connected by $\boldsymbol{\lambda}$, we first orthogonalize the network using Eq.~\eqref{eq:orth} with mapping tensors $\mathcal{R}_A$ and $\mathcal{R}_B$:
\begin{equation}\label{eq:orth_network}
  \begin{aligned}
    \relax[\mathcal{A}_\perp]^{\mathbf{q}_A}_{\mathbf{c}^{-}_A,j} &= \sum_i [\mathcal{A}]^{\mathbf{q}_A}_{\mathbf{c}^{-}_A,i}(\mathcal{R}_A)_{ij}, \\
    \relax[\mathcal{B}_\perp]^{\mathbf{q}_B}_{\mathbf{c}^{-}_B,j} &= \sum_i [\mathcal{B}]^{\mathbf{q}_B}_{\mathbf{c}^{-}_B,i}(\mathcal{R}_B)_{ij}, \\
    (\boldsymbol{\lambda}_\perp)_{il} &= \sum_{j,k} (\mathcal{R}_A^+)_{ij} \boldsymbol{\lambda}_{jk} (\mathcal{R}_B^+)_{lk}.
  \end{aligned}
\end{equation}
where $\mathbf{c}^{-}_A$ and $\mathbf{c}^{-}_B$ denote the classical indices not attached to $\boldsymbol{\lambda}$.
The orthogonalized sites $\mathcal{A}_\perp$ and $\mathcal{B}_\perp$ have effective rank determined by their Gram matrices.

Because $\boldsymbol{\lambda}_\perp$ is classical, it admits an SVD, as shown in Fig.~\ref{fig:orth}(b).
Theorem~2 identifies this SVD with the Schmidt decomposition across the cut.

{\bf Theorem 2}.
Let $[\mathcal{A}_\perp]_{\mathbf{c}_A}^{\mathbf{q}_A}$ and $[\mathcal{B}_\perp]_{\mathbf{c}_B}^{\mathbf{q}_B}$ be hybrid sites linked by a classical connector $\boldsymbol{\lambda}_\perp$ and orthogonalized on the shared indices as in Eq.~\eqref{eq:orth}.
Merge the remaining indices of $[\mathcal{A}_\perp]$ into a composite label $\alpha=(\mathbf{c}^{-}_A,\mathbf{q}_A)$ and those of $[\mathcal{B}_\perp]$ into $\beta=(\mathbf{c}^{-}_B,\mathbf{q}_B)$.
The amplitude matrices are
\begin{equation}\label{eq:amp_AB}
  \begin{aligned}
    A_{\alpha k} &= [\mathcal{A}_\perp]^{\mathbf{q}_A}_{\mathbf{c}^{-}_A,k}=\langle\mathbf{q}_A|\psi_{\mathbf{c}^{-}_A,k}\rangle,\\
    B_{l\beta} &= [\mathcal{B}_\perp]^{\mathbf{q}_B}_{\mathbf{c}^{-}_B,l}=\langle\mathbf{q}_B|\phi_{\mathbf{c}^{-}_B,l}\rangle.
  \end{aligned}
\end{equation}
The equivalent bipartite amplitude matrix is
\begin{equation}\label{eq:amp_matrix}
  \Gamma_{\alpha\beta}=\sum_{k,l}A_{\alpha k}(\boldsymbol{\lambda}_\perp)_{kl}B_{l\beta}.
\end{equation}
For the bipartite state $|\Psi\rangle=\sum_{\alpha\beta}\Gamma_{\alpha\beta}|\alpha\rangle|\beta\rangle$, the nonzero singular values of $\boldsymbol{\lambda}_\perp$ equal the Schmidt coefficients of $|\Psi\rangle$.

{\em Proof}.
In matrix form, $\Gamma=A\boldsymbol{\lambda}_\perp B$.
Orthogonalization gives $A^\dagger A=BB^\dagger=I$.
Hence
\begin{equation}\label{eq:GammaGammadag}
  \Gamma\Gamma^\dagger = A\boldsymbol{\lambda}_\perp B B^\dagger \boldsymbol{\lambda}_\perp^\dagger A^\dagger = A\boldsymbol{\lambda}_\perp \boldsymbol{\lambda}_\perp^\dagger A^\dagger.
\end{equation}
Because $A$ has orthonormal columns, $A\boldsymbol{\lambda}_\perp \boldsymbol{\lambda}_\perp^\dagger A^\dagger$ and $\boldsymbol{\lambda}_\perp \boldsymbol{\lambda}_\perp^\dagger$ share the same nonzero eigenvalues.
Thus $\Gamma$ and $\boldsymbol{\lambda}_\perp$ have the same nonzero singular values, which are the Schmidt coefficients of $|\Psi\rangle$. $\blacksquare$

As in conventional tensor-network algorithms, the connector admits an SVD $\boldsymbol{\lambda}_\perp=USV$.
Truncation keeps the leading singular values and the matching blocks of $U$ and $V$.
These blocks are absorbed into the hybrid sites in opposite index order, as shown in Fig.~\ref{fig:orth}(b).
\begin{equation}\label{eq:svd_network}
  \begin{aligned}
    \relax[\widetilde{\mathcal{A}}]^{\mathbf{q}_A}_{\mathbf{c}^{-}_A,l} &= \sum_{i} [\mathcal{A}]^{\mathbf{q}_A}_{\mathbf{c}^{-}_A,i}(\mathcal{R}_A U)_{il}, \\
    \relax[\widetilde{\mathcal{B}}]^{\mathbf{q}_B}_{\mathbf{c}^{-}_B,l} &= \sum_{i} [\mathcal{B}]^{\mathbf{q}_B}_{\mathbf{c}^{-}_B,i}(\mathcal{R}_B V^T)_{il}.
  \end{aligned}
\end{equation}
They remain linked by the diagonal connector $S$, whose entries $S_{ll}=s_l$ are the retained Schmidt coefficients.
After retaining the largest $\chi$ singular values, we renormalize them so that $\sum_l s_l^{2}=1$.
The connector index on $[\widetilde{\mathcal{A}}]$ and $[\widetilde{\mathcal{B}}]$ then has size $\chi\le d$, where $d$ is the cut dimension before truncation.
Because $U$ and $V$ are truncated isometries, the updated sites remain normalized along the truncated connector and the weights stay at unity.

Each hybrid site may still encode high entanglement within its physical qubits, including volume-law entanglement local to a subsystem, while the bond dimension $\chi$ limits only the Schmidt modes across the cut.
This separation matches partitioned Trotter dynamics, in which remote gates increase entanglement between subsystems but leave intra-subsystem correlations to be carried by the physical states at each site.
Across connector topologies, the cut truncation plays the same role as bond truncation in classical tensor networks.
On tree THTNs, truncating each edge controls the Schmidt coefficients between the corresponding regions.
On loopy or higher-dimensional layouts, the method inherits the familiar accuracy and cost limitations of classical TN algorithms.

\subsection{Sampling-based observable expectations}

The expectation of a local observable, such as a spin component or an energy density, is obtained by contracting the THTN representation of $|\Psi(t)\rangle$ with its conjugate.
Since any qubit observable admits a finite decomposition into local Pauli strings, we consider local Pauli observables without loss of generality.

For a system of $M$ subsystems whose hybrid sites have connector dimension $\chi$, the global expectation can be estimated by sampling those connector indices rather than summing them explicitly.
We illustrate the two-site case.
Hybrid sites $[\widetilde{\mathcal{A}}]_{\mathbf{c}_A}^{\mathbf{q}_A}$ and $[\widetilde{\mathcal{B}}]_{\mathbf{c}_B}^{\mathbf{q}_B}$ on subsystems $A$ and $B$ are linked by the diagonal connector $S$, with $S_{ll}=s_l\ge 0$.
Writing $[\widetilde{\mathcal{A}}]^{\mathbf{q}_A}_{\mathbf{c}^{-}_A,l}=|\psi_{\mathbf{c}^{-}_A,l}\rangle$ and $[\widetilde{\mathcal{B}}]^{\mathbf{q}_B}_{\mathbf{c}^{-}_B,l}=|\phi_{\mathbf{c}^{-}_B,l}\rangle$, the expectation indexed by the classical indices not attached to the connector is calculated by
\begin{equation}\label{eq:global_exp_two_cite}
  \begin{aligned}
    &[\langle\Psi_{AB}|\hat{O}|\Psi_{AB}\rangle]_{\mathbf{c}^{-}_A,\mathbf{c}'^{-}_A,\mathbf{c}^{-}_B,\mathbf{c}'^{-}_B} \\
    =& \sum_{l,l',\mathbf{q}_A,\mathbf{q}'_A,\mathbf{q}_B,\mathbf{q}'_B} S_{ll}S_{l'l'}[\widetilde{\mathcal{A}}^*]^{\mathbf{q}_A}_{\mathbf{c}^{-}_A,l}[\hat{O}_A]^{\mathbf{q}_A,\mathbf{q}'_A}[\widetilde{\mathcal{A}}]^{\mathbf{q}'_A}_{\mathbf{c}'^{-}_A,l'}\\
    &\times[\widetilde{\mathcal{B}}^*]^{\mathbf{q}_B}_{\mathbf{c}^{-}_B,l}[\hat{O}_B]^{\mathbf{q}_B,\mathbf{q}'_B}[\widetilde{\mathcal{B}}]^{\mathbf{q}'_B}_{\mathbf{c}'^{-}_B,l'} \\
    =&\sum_{l,l'} s_l s_{l'}\langle\psi_{\mathbf{c}^{-}_A,l}|\hat{O}_A|\psi_{\mathbf{c}'^{-}_A,l'}\rangle\langle\phi_{\mathbf{c}^{-}_B,l}|\hat{O}_B|\phi_{\mathbf{c}'^{-}_B,l'}\rangle \\
    =&\sum_{l,l'} p_l p_{l'}\,\mathcal{Z}\,\mathcal{Z}\,\langle\psi_{\mathbf{c}^{-}_A,l}|\hat{O}_A|\psi_{\mathbf{c}'^{-}_A,l'}\rangle\langle\phi_{\mathbf{c}^{-}_B,l}|\hat{O}_B|\phi_{\mathbf{c}'^{-}_B,l'}\rangle,
  \end{aligned}
\end{equation}
where $\mathcal{Z}=\sum_l s_l$ and $p_l=s_l/\mathcal{Z}$.
Because $s_l\ge 0$, the Schmidt coefficients define a probability distribution $\{p_l\}$.
On hardware, each experimental shot contributes to a running average.
One first samples the classical connector indices independently from $\{p_l\}$ (here $l$ and $l'$), prepares the corresponding hybrid-site branches, and obtains local measurement outcomes $\widehat{o}_A$ and $\widehat{o}_B$.
The accumulated estimate on each shot carries the product of the reweighting factors $\mathcal{Z}$ from that shot's draws.
For the two-site bond the contribution is $\mathcal{Z}\,\mathcal{Z}\,\widehat{o}_A\widehat{o}_B$, which is an unbiased estimator of the bond sum in Eq.~\eqref{eq:global_exp_two_cite}.

Unlike quasi-probability sampling, here the estimate is directly sampled from a true probability distribution.
The sampling overhead is assessed by the variance of this shot estimator.
For one fixed element of the two-site bond with free classical indices, one has $X=\mathcal{Z}^{2}\,\widehat{o}_A\widehat{o}_B$.
For local Pauli observables the outcomes satisfy $|\widehat{o}_A\widehat{o}_B|\le 1$, so as a worst-case bound $\mathrm{Var}(X)\le\mathbb{E}[|X|^{2}]\le\mathcal{Z}^{4}$, and $O(\mathcal{Z}^{4}/\varepsilon^{2})$ shots suffice for accuracy $\varepsilon$ on that element.
The renormalization after the truncation ensures $\sum_l s_l^{2}=1$, so $\mathcal{Z}\le\sqrt{\chi}$ by the Cauchy--Schwarz inequality.
The shot cost of this estimator is then $O(\chi^{2}/\varepsilon^{2})$ per element.
This bound applies to readout of a network already truncated to $\chi$.
With $\chi$ fixed, that readout cost no longer depends on the number of remote gates absorbed into the connector.

For a contraction of the full network one may draw every connector index in a single shot and accumulate the product of reweighting factors.
After truncation that protocol scales as $O(\chi^{2B}/\varepsilon^{2})$ for $B$ cuts.
The shot cost is exponential in $B$, but polynomial in $\chi$ once $B$ is fixed.
For quantum simulation with few hybrid sites, $B$ stays small, so joint sampling scales as a polynomial in $\chi$.
Circuit knitting has no analogous truncated-$\chi$ scaling. Its overhead is a product over cuts.
For a remote Pauli rotation $e^{-i\theta_{ij}\sigma_i\otimes \sigma_j}$ the single-cut factor is $(1+2|\sin\theta_{ij}|)^{2}$.
The product remains exponential in the cut count for any number of sites, even when each factor is modest.
Details are given in the Supplemental Material.

\subsection{From Trotter simulation to distributed quantum computation}\label{sec:remark}

The primary application developed above is Trotter-based simulation on a partitioned lattice.
After Trotter splitting~\cite{Suzuki1991,Childs2021}, the evolution is a product of local and two-body operations $e^{-iH\Delta t}$.
Local terms update hybrid sites.
Boundary-spanning terms are inserted through Eq.~\eqref{eq:apply}.
After each remote gate, the affected cut is truncated via Eqs.~\eqref{eq:orth_network} and~\eqref{eq:svd_network}.
The interface bond dimension is thereby kept at most $\chi$.
Observables follow from the truncated network through Eq.~\eqref{eq:global_exp_two_cite}.

The same remote-gate representation is not limited to Trotter dynamics.
Controlled-Pauli gates admit rank-$2$ decompositions~\cite{Bravyi2016}, and can be written as
\begin{equation}
  \mathrm{C}\text{-}P=e^{i\frac{\pi}{4}}e^{-i\frac{\pi}{4}Z_1}e^{-i\frac{\pi}{4}P_2}e^{i\frac{\pi}{4}Z_1\otimes P_2}
\end{equation}
where $P$ is a single-qubit Pauli operator and the subscripts label the qubits on which the operators act.
Any other two-qubit unitary can be reduced to Pauli-string form by local rotations (KAK decomposition)~\cite{Khaneja2001}.
For generic partitioned circuits, the cut entanglement need not be low rank, and a small retained $\chi$ may incur a large truncation error.
Without truncation the THTN readout samples cut indices with overhead $O(4^{n_{\rm cut}})$ for rank-$2$ CNOT decompositions, matching the scaling of circuit knitting~\cite{Piveteau2024,Mitarai2021} while remaining distinct in mechanism.
In Trotter dynamics of systems with bounded correlations, the Schmidt spectrum across a cut is often concentrated on a few large singular values, so $\mathcal{Z}$ remains moderate after truncation.
In more generic computations, that spectrum is typically flatter, which enlarges $\mathcal{Z}$ and raises the sampling overhead.

Even when truncation does not compress the cut, the present framework still supports modular execution and can reduce long-range routing on each device.
The same framework applies not only to remote gates on physical qubits, but also to operations between logical qubits.

\section{Numerical experiments}\label{sec:numerical}

\subsection{Numerical methods}

\begin{figure}[t]
  \centering
  \includegraphics[width=0.48\textwidth]{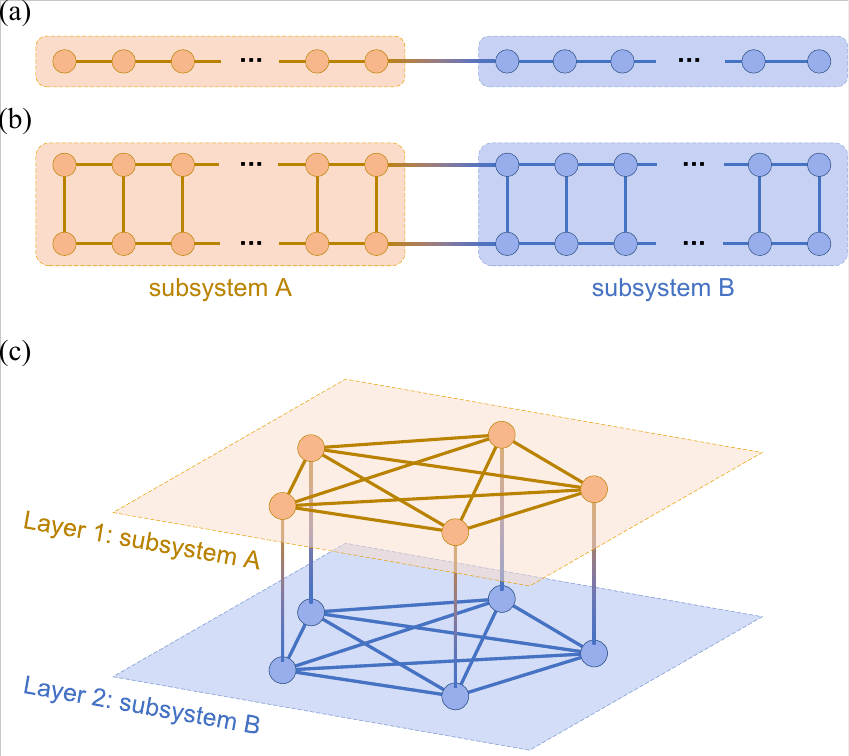}
  \caption{Lattice partitions for the numerical benchmarks.
  (a)~One-dimensional chain with a midpoint cut.
  (b)~Two-leg ladder with a rung cut at the middle of the ladder.
  (c)~Layered systems with one hybrid site per layer.
  }\label{fig:Q1Dsys}
\end{figure}

The THTN framework is a distributed protocol for quantum simulation on modest devices, not a new classical numerical method.
We nevertheless validate it by classical simulation of partitioned Trotter evolution on one-dimensional, quasi-one-dimensional, and layered spin models.
Chains and ladders serve as successive checks.
The layered partition further probes truncation at the cut under intra- and interlayer scale separation.
The lattice is partitioned into two subsystems by a fixed spatial cut, as shown in Fig.~\ref{fig:Q1Dsys}.
We use second-order Suzuki--Trotter splitting~\cite{Suzuki1991,Childs2021}.
Connectors are updated and truncated after each remote Trotter gate is applied.
Connector SVD retains the largest $\chi$ Schmidt coefficients under a fixed bond-dimension limitation.
Hybrid-site states are evolved with a classical MPS-based simulator~\cite{Vidal2003,Liu2020,GuoLiu2019}.
Observable expectations are obtained by tensor network contraction, rather than connector sampling.
All runs start from random product initial states.
The baseline is TEBD on a one-dimensional MPS chain~\cite{Vidal2004,Schollwock2011}, the representative method for unitary dynamics of local one-dimensional Hamiltonians.
The structural difference is that TEBD truncates internal MPS bonds, while THTN only truncates the classical connector on the partition cut.
Other settings, including the Hamiltonian, initial states, and step sizes, are kept the same.

Benchmarks on one-dimensional chains use $N=10$, $14$, $20$, and $30$.
For all cases, we use $\Delta t = 0.05$.
Each chain uses $30$ Trotter steps ($20$ steps for $N = 30$).
For $N=10$ and $14$ in all geometries, we tested TEBD at $\chi=4,8,\ldots,64$ and THTN at $\chi=4,8,16$, and we also computed exact results.
For $N=20$ and $30$, TEBD extends to $\chi=4,8,\ldots,128$ and THTN to $\chi=16$.
Each system size is simulated in three independent runs with random product initial states.
The primary comparison between THTN and TEBD is restricted to $\chi=4,8,16$, where both methods are reported.
TEBD results at larger $\chi$ are used only to show how large a bond dimension TEBD needs to approach the comparable accuracy.
The quasi-one-dimensional and layered geometries are reported only at $N=10$ and $14$, because at larger system sizes TEBD at $\chi=128$ may not be a reliable reference.

\begin{figure*}[t]
  \centering
  \includegraphics[width=\textwidth]{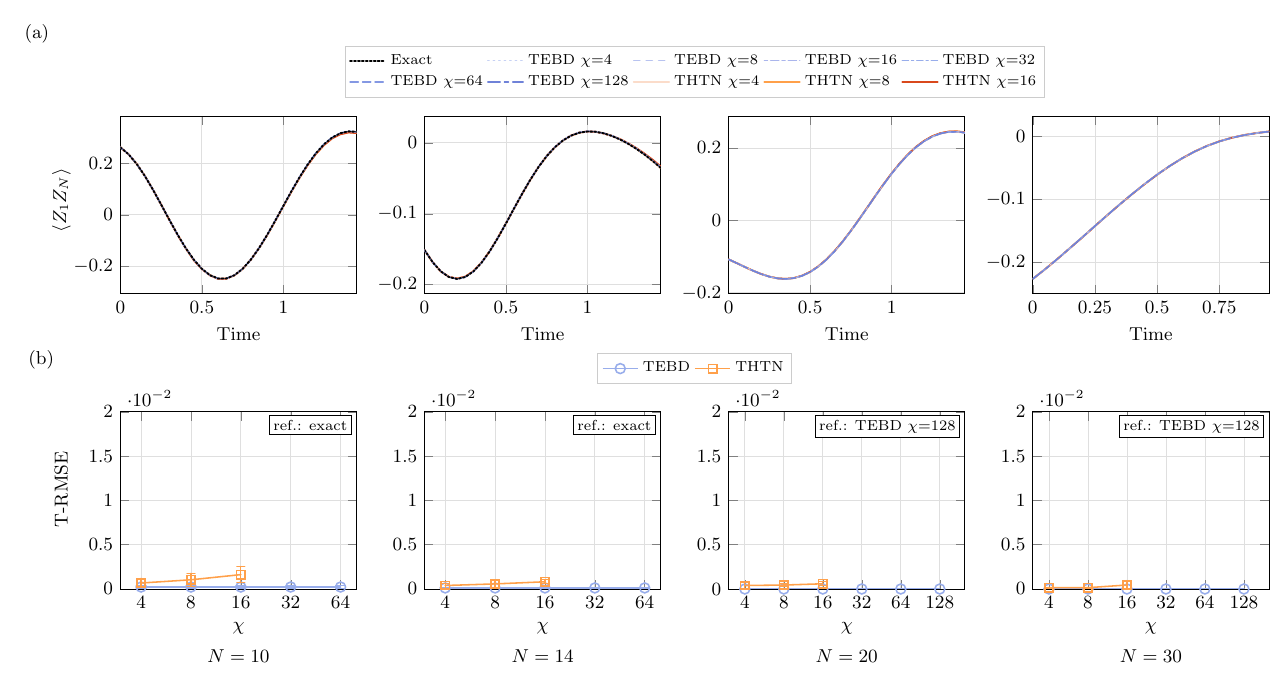}
  \caption{Transverse-field Ising model (TFIM) under second-order Trotter evolution.
  Each system size is simulated in three independent runs with random product initial states.
  (a) Time evolution of the end-to-end correlator $\langle Z_1 Z_N\rangle$ for $t=\mathrm{step}\,\Delta t$, shown for one representative run.
  (b) Mean T-RMSE over those runs for every curve shown in panel~(a), including the reference, with error bars denoting one sample standard deviation.
  The reference is exact evolution at $N=10,14$ and TEBD with $\chi=128$ at $N=20,30$.
  At matched $\chi$, both methods stay close to the reference.
  }\label{fig:Ising}
\end{figure*}

For $N=10$ and $14$, exact results are the reference. For $N=20$ and $30$, the results of TEBD at $\chi=128$ serve as the reference.
Deviation of an observable from its reference over time is quantified by the trajectory root-mean-square error (T-RMSE).
T-RMSE in Figs.~\ref{fig:Ising} to~\ref{fig:LayeredIsing} is the mean over these runs with error bars of one sample standard deviation, while panel~(a) of each figure shows one representative run.

\subsection{One-dimensional systems}\label{sec:1d}

On one-dimensional chains we test the transverse-field Ising and Heisenberg XXZ models.
Sites are numbered left to right as in Fig.~\ref{fig:Q1Dsys}(a), with $N/2$ sites in each subsystem and the midpoint cut between qubits $N/2$ and $N/2+1$.
On this geometry, the TFIM is simple and remains within the area-law regime at the sizes considered here, so it serves as a correctness check of the THTN update and truncation loop.
The XXZ model activates more involved two-body Trotter terms across the cut.
It is used to compare the accuracy of THTN and TEBD at matched $\chi$ when entanglement across the cut grows rapidly.

\subsubsection{Transverse-field Ising model}

The Hamiltonian is
\begin{equation}\label{eq:tfim}
  H_{\mathrm{Ising}}=\sum_{i=1}^{N-1}J_{i}\sigma_i^x\sigma_{i+1}^x + \sum_{i=1}^{N} h_z \sigma_i^z,
\end{equation}
where $N$ is the chain length, $h_z=0.5$ is the transverse-field strength, and $J_i$ is the coupling strength on bond $(i,i+1)$, set to $J_i=1$ on odd bonds and $J_i=0.25$ on even bonds.

We evolve chains with $N=10$ to $30$ from random product initial states and monitor the end-to-end correlator $\langle Z_1 Z_N\rangle$.
Figure~\ref{fig:Ising} shows the THTN and TEBD trajectories together with T-RMSE.
At matched $\chi$, the THTN, TEBD, and reference curves overlap across these sizes.

The THTN T-RMSE nevertheless lies slightly above that of TEBD at matched $\chi$.
Each THTN remote-gate update orthogonalizes the hybrid sites through a Gram matrix, truncates the classical connector by SVD, and absorbs the mapping tensors back into the sites.
These extra steps are unnecessary in a standard TEBD update on a single MPS. They can introduce a small extra numerical error even when $\chi$ already keeps the Schmidt spectrum across the cut.

\subsubsection{Heisenberg XXZ model}

\begin{figure*}
  \centering
  \includegraphics[width=\textwidth]{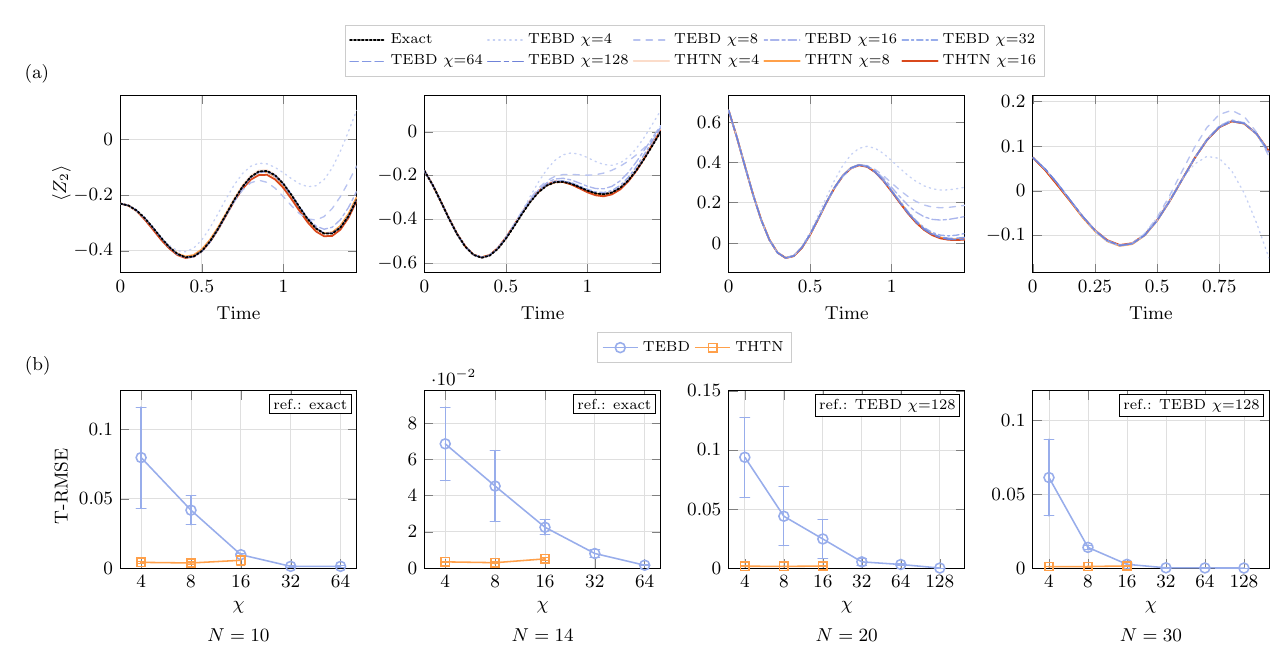}
  \caption{Heisenberg XXZ chain under second-order Trotter evolution.
  Each system size is simulated in three independent runs with random product initial states.
  (a) Time evolution of the local expectation $\langle Z_2\rangle$ for $t=\mathrm{step}\,\Delta t$, shown for one representative run.
  (b) Mean T-RMSE over those runs for every curve shown in panel~(a), including the reference, with error bars denoting one sample standard deviation.
  The reference is exact evolution at $N=10,14$ and TEBD with $\chi=128$ at $N=20,30$.
  }\label{fig:Heisenberg}
\end{figure*}

The Ising benchmarks above mainly check correctness when cross-cut correlations stay modest.
We next turn to the Heisenberg XXZ chain, where each remote Trotter term activates $XX$, $YY$, and $ZZ$ terms.
The interface entanglement grows faster than in the Ising case.
At matched $\chi$, the comparison therefore tests whether truncating Schmidt modes on the cut yields higher accuracy than truncating internal MPS bonds.

The Hamiltonian is
\begin{equation}
  H_{\mathrm{XXZ}}=\sum_{i=1}^{N-1}(J\sigma_i^x\sigma_{i+1}^x+J\sigma_i^y\sigma_{i+1}^y + \Delta\sigma_i^z\sigma_{i+1}^z) + \sum_{i=1}^{N} h_z \sigma_i^z,
\end{equation}
where $N$ is the chain length, $J=1$ is the coupling strength, $\Delta=0.5$ is the anisotropy parameter of the $ZZ$ interaction relative to $J$, and $h_z=1$ is the longitudinal-field strength.

We evolve chains with $N=10$ to $30$ and monitor $\langle Z_2\rangle$.
Figure~\ref{fig:Heisenberg} shows the THTN and TEBD trajectories together with T-RMSE.
In panel~(a), TEBD at small $\chi$ departs from the reference within the first few steps, whereas THTN at the same $\chi$ stays close to the exact reference or to TEBD at large $\chi$.
For $N=20$ and $30$, T-RMSE is measured against TEBD at $\chi=128$, and TEBD at low bond dimension fluctuates strongly across independent runs.

Within $\chi=4$ to $16$, the TEBD T-RMSE is typically of order $10^{-2}$, while THTN stays near $10^{-3}$.
TEBD reaches order $10^{-3}$ only near $\chi=64$, and for $N=20$ and $30$ it meets the reference floor at $\chi=128$.
Thus THTN at a moderate cut bond dimension already suffices where TEBD needs a much larger internal bond dimension.

\subsection{Quasi-one-dimensional systems}\label{sec:ladder}

Beyond a single chain, we next consider a quasi-one-dimensional geometry.
We take the two-leg ladder in Fig.~\ref{fig:Q1Dsys}(b) as a representative case.
Qubits in subsystem~A are numbered before those in~B.
In~A, sites run left to right along the first leg ($1,\ldots,\ell_1$), where $\ell_1=\lceil N/4\rceil$, then continue along the second leg.
Subsystem~B follows the same pattern, with leg length $\ell_2=N/2-\ell_1$.
The rung cut pairs $(\ell_1,\,2\ell_1+1)$ and $(2\ell_1,\,2\ell_1+\ell_2+1)$, and the inter-block bonds share one classical connector after the merge.
Therefore, the THTN layout stays one-dimensional.
We test whether the same advantage of truncation at the cut persists when the interface is a rung rather than a chain midpoint.

\begin{figure}
  \centering
  \includegraphics[width=\columnwidth]{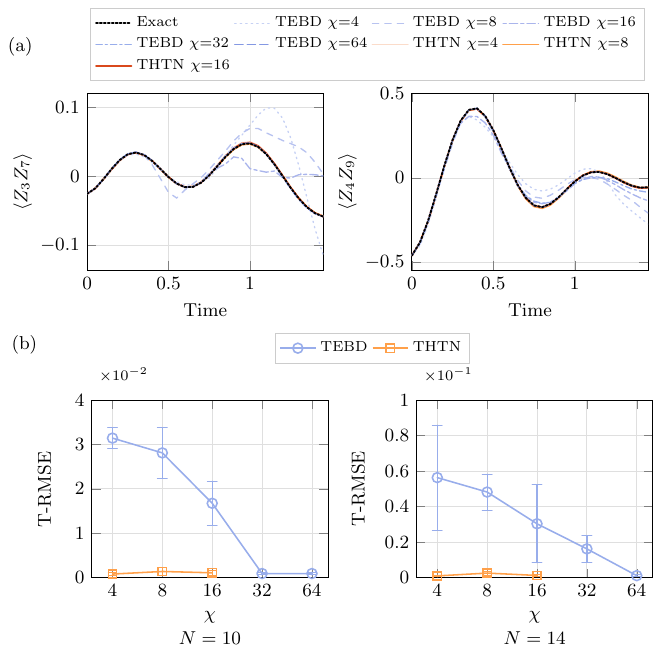}
  \caption{Two-leg ladder Ising model at $N=10$ and $14$, with intra-block $J=1$, inter-block $J=0.25$, $h_z=0.5$, and $\Delta t=0.05$.
  Each system size is simulated in three independent runs with random product initial states.
  (a) Time evolution of the correlators $\langle Z_3 Z_7\rangle$ and $\langle Z_4 Z_9\rangle$ across the rung cut for $t=\mathrm{step}\,\Delta t$, shown for one representative run.
  (b) Mean T-RMSE over those runs for every curve shown in panel~(a), including the reference, with error bars denoting one sample standard deviation.
  }\label{fig:TwoLegLadder}
\end{figure}

We simulate a transverse-field Ising model of the same interaction type as Eq.~\eqref{eq:tfim}, now on the ladder bonds.
The intra-block coupling strength is $J=1$, the inter-block coupling strength is $J=0.25$, and the transverse-field strength is $h_z=0.5$.
The observable is the correlator across the rung cut, $\langle Z_3 Z_7\rangle$ at $N=10$ and $\langle Z_4 Z_9\rangle$ at $N=14$.
Figure~\ref{fig:TwoLegLadder} shows the trajectories and T-RMSE.
Within $\chi=4$ to $16$, TEBD remains of order $10^{-2}$ in T-RMSE, while THTN stays near $10^{-3}$.
TEBD reaches order $10^{-3}$ only at $\chi=64$.
Truncation at the cut therefore remains advantageous on this rung cut.

\subsection{Layered systems}\label{sec:layered}

\begin{figure}
  \centering
  \includegraphics[width=\columnwidth]{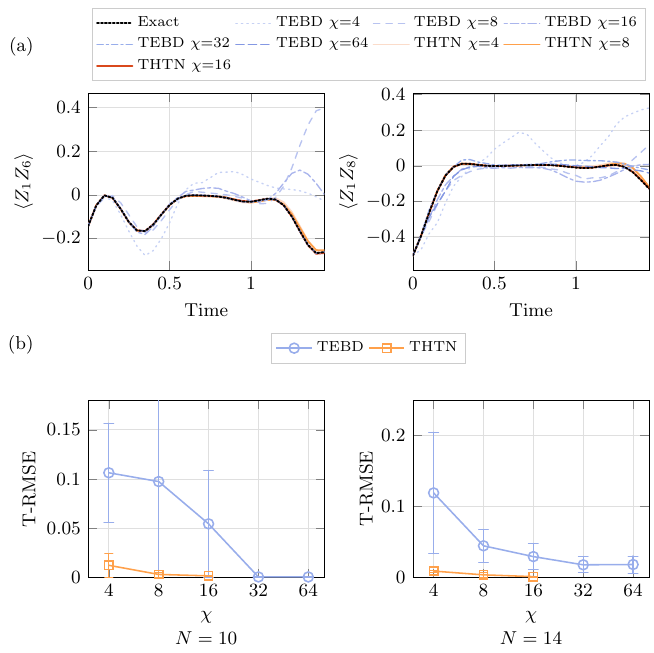}
  \caption{Layered fully connected Ising model at $N=10$ and $14$, with intra-layer $J=1$, interlayer $J=0.25$, $h_z=0.5$, and $\Delta t=0.05$.
  Each system size is simulated in three independent runs with random product initial states. Each layer holds $N/2$ sites.
  (a) Time evolution for $t=\mathrm{step}\,\Delta t$ of the interlayer correlator $\langle Z_1 Z_{N/2+1}\rangle$, shown for one representative run.
  (b) Mean T-RMSE over those runs for every curve shown in panel~(a), including the reference, with error bars denoting one sample standard deviation.
  }\label{fig:LayeredIsing}
\end{figure}

We next move from this quasi-one-dimensional setting to a layered geometry, shown in Fig.~\ref{fig:Q1Dsys}(c).
Each layer is assigned to one subsystem with fully connected sites.
Qubits in~A precede those in~B, with the same within-layer ordering in both layers.
The interlayer cut pairs site~$i$ with $i+N/2$ for $i\in\{1,\ldots,N/2\}$.
Weaker interlayer bonds are applied as remote gates on the connector.
The same strong-intra / weak-inter pattern appears in van der Waals heterostructures (e.g., bilayer graphene)~\cite{Geim2013,McCann2013,Kennes2021,Cao2018}, and is shared by bilayer spin and Hubbard prototypes used in quantum simulation.
Here we take a minimal layered Ising instance with that scale separation.
In this layout the physical model is no longer one-dimensional, yet the THTN representation keeps a single classical connector between the two layers.
The benchmark therefore tests whether evolving strong intra-layer correlations within the hybrid sites, while truncating only the weaker interlayer cut, favors THTN over TEBD at matched $\chi$.

We likewise use a transverse-field Ising model of the same interaction type as Eq.~\eqref{eq:tfim}, with couplings on the intra- and interlayer edges.
The intra-layer coupling strength is $J=1$, the interlayer coupling strength is $J=0.25$, and the transverse-field strength is $h_z=0.5$.
We monitor the interlayer correlator $\langle Z_1 Z_{N/2+1}\rangle$.
Figure~\ref{fig:LayeredIsing} shows the trajectories and T-RMSE.
Within $\chi=4$ to $16$, the TEBD T-RMSE falls from order $10^{-1}$ to about $10^{-2}$, while THTN stays near $10^{-3}$.
At $\chi=64$, TEBD reaches order $10^{-4}$ at $N=10$, but remains of order $10^{-2}$ at $N=14$.
At $N=10$, with five sites per layer, the maximum Schmidt rank across the cut is $2^{5}=32$, and TEBD at $\chi\ge 32$ saturates this MPS capacity.
Thus at $N=10$ TEBD catches up only at large internal bond dimension, whereas at $N=14$ even $\chi=64$ does not close the gap to THTN.
The layered cut therefore shows the clearest advantage of evolving strong intra-layer correlations within the hybrid sites and truncating only the weaker interlayer connector.

\section{Discussion}\label{sec:discussion}

Partitioned digital simulation on distributed registers must keep remote two-body gates under control as the Trotter depth grows.
We have introduced a truncated hybrid tensor network (THTN) for that setting.
Interface truncation retains the leading Schmidt modes across each cut up to bond dimension $\chi$, while entanglement within each subsystem need not be compressed.
Relative to prior hybrid-tensor constructions, the present work supplies a dynamical update under successive remote Trotter gates, a truncation at the partition cut based on Gram orthonormalization~\cite{Schuhmacher2024} with recovery through a possibly rank-deficient mapping tensor $\mathcal{R}$, and a readout that samples non-negative Schmidt coefficients rather than signed quasi-probabilities.

The numerical benchmarks support this picture of truncation at the cut.
At equal $\chi$, THTN tracks exact references or references at large $\chi$ more closely than TEBD when cross-cut correlations are not near area-law.
The gain is visible on the XXZ chain and the quasi-one-dimensional ladder, and clearest on the layered geometry with weak interlayer bonds.
These tests cover chain, ladder, and layered partitions.
THTN can support non-tree connector topologies.
In practice, however, the same framework can cast non-one-dimensional models onto a one-dimensional cut, as shown by the layered-partition results.
Strong intra-layer correlations evolve within the hybrid sites.
Only the weaker cross-cut bonds are truncated on the classical connectors.
Layered partitions with that scale separation are a natural target for this picture of truncation at the cut, as in van der Waals heterostructures~\cite{Geim2013,McCann2013,Kennes2021,Cao2018}.

For experimental demonstration, with the numerical evidence in place, the next step is to realize the truncation loop on physical devices by streamlining Gram estimation from hybrid-site overlaps and the absorption of the truncated bond into the sites under realistic noise.
Both steps act locally on each register rather than requiring a global reconstruction, and may therefore be accessible targets for near-term demonstration.
On near-term devices a complete demonstration still needs simplified experimental protocols before large-scale Trotter runs.
Designing hybrid contraction algorithms may further improve overall efficiency.
Encoding connector labels on a quantum interconnect further reduces the classical resource demand and can improve overall efficiency when $\chi$ stays moderate.
Such optimizations can be folded into modular compilation by choosing cuts where interface correlations remain tractable~\cite{Harrow2025,Cuomo2023}.
Taken together, THTN offers a route to distributed quantum simulation by truncating at the cut when interface correlations remain manageable.

\begin{acknowledgments}

We appreciate the helpful discussions with other members of the QUANTA group.
This work was funded by the National Key R\&D Program of China under Grant No. 2024YFB4504001, the National Natural Science Foundation of China under Grant No. 62401572, the Innovation Research Foundation of National University of Defense Technology, the Aid Program for Science and Technology Innovative Research Team in Higher Educational Institutions of Hunan Province, and the Fundamental and Interdisciplinary Disciplines Breakthrough Plan of the Ministry of Education of China under Grant No. JYB2025XDXM202.

\end{acknowledgments}

\clearpage
\onecolumngrid

\setcounter{section}{0}
\setcounter{equation}{0}
\setcounter{figure}{0}
\setcounter{table}{0}
\renewcommand{\theequation}{S\arabic{equation}}
\renewcommand{\thefigure}{S\arabic{figure}}
\renewcommand{\thetable}{S\Roman{table}}
\renewcommand{\bibnumfmt}[1]{[S#1]}
\renewcommand{\citenumfont}[1]{S#1}

\begin{center}\bf\large
    Supplemental Material: Truncated hybrid tensor networks for distributed quantum simulation
\end{center}

\section{Cases of contracting hybrid tensors}

Contracting hybrid tensors follows the same index-matching logic as in classical tensor computation, except that each contracted index may be classical or quantum.
Each element of the result can be either a classical scalar or a quantum state, depending on whether all quantum indices are contracted.
Hybrid sites are denoted $[\mathcal{A}]_{\mathbf{c}}^{\mathbf{q}}=|\psi_{\mathbf{c}}\rangle$ as in the main text, where the subscripts $\mathbf{c}=\{c_1,c_2,\ldots\}$ are classical indices that specify the site shape and each state contains qubits labeled by the superscripts $\mathbf{q}=\{q_1,q_2,\ldots\}$.
Below we enumerate the ten contraction cases according to the tensor types and the index types involved.
For clarity we first treat a single contracted index.
Extensions to multiple indices follow the same pattern with additional ancilla qubits and multi-controlled gates.

Throughout Cases~2 to~10, the circuit diagrams and evolution equations show only the key steps of the state evolution.
The intermediate kets written in those equations are not necessarily normalized.
The resulting physical state is normalized, and its scale relative to the unnormalized contraction $|\gamma\rangle$ is the norm $\mathcal{N}=\sqrt{\langle\gamma|\gamma\rangle}$.

\subsection{Case 1: a classical tensor contracted with a classical tensor}

This case is identical to conventional tensor contraction and produces a classical tensor.

\subsection{Case 2: a classical tensor contracted with a hybrid tensor along a classical index}

Consider a classical tensor as a vector $a_i$ and a hybrid tensor $[\mathcal{B}]_{ik}=w_{ik}|\beta_{ik}\rangle$.
Contracting along the classical index $i$ gives
\begin{equation}\label{eq:case2_gamma}
|\gamma_k\rangle = \sum_{i} a_iw_{ik} |\beta_{ik}\rangle.
\end{equation}

\begin{figure}[h]
  \centering
  \includegraphics[width=0.5\linewidth]{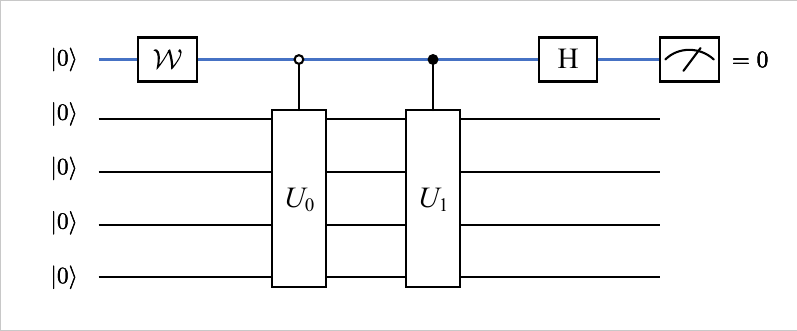}
  \caption{Quantum circuit for contracting a classical tensor with a hybrid tensor along a classical index (Case~2).
  An ancilla in superposition selects the hybrid branches, and post-selection on $|0\rangle$ yields the contracted hybrid state.}
  \label{fig:case2}
\end{figure}

The corresponding quantum circuit is shown in Fig.~\ref{fig:case2}.
For each fixed $k$, unitaries $U_0$ and $U_1$ satisfy $U_0|0\rangle=|\beta_{0k}\rangle$ and $U_1|0\rangle=|\beta_{1k}\rangle$, and a $\mathcal{W}$ gate on the ancilla qubit labeled $c$ prepares $a_0w_{0k}|0\rangle_c+a_1w_{1k}|1\rangle_c$.

The state evolution (key steps only) is
\begin{equation}
\begin{aligned}
|0\rangle_c|0\rangle
& \xrightarrow{\mathcal{W}} (a_0w_{0k}|0\rangle_c+a_1w_{1k}|1\rangle_c)|0\rangle \\
& \xrightarrow{\mathrm{C}\text{-}U} a_0w_{0k}|0\rangle_c|\beta_{0k}\rangle + a_1w_{1k}|1\rangle_c|\beta_{1k}\rangle \\
& \xrightarrow[M_c=0]{H} a_0w_{0k}|\beta_{0k}\rangle + a_1w_{1k}|\beta_{1k}\rangle,
\end{aligned}
\end{equation}
where $M_c$ means the measurement on the ancilla (the qubit represented by the blue line).
Post-selection returns a normalized physical state that differs from Eq.~\eqref{eq:case2_gamma} by the norm $\mathcal{N}_k=\sqrt{\langle\gamma_k|\gamma_k\rangle}$, with
\begin{equation}
  \mathcal{N}_k=\sqrt{\sum_{ii'}a_ia_{i'}^*w_{ik}w_{i'k}^*\langle\beta_{i'k}|\beta_{ik}\rangle}.
\end{equation}
Whenever the branch overlaps $\langle\beta_{i'k}|\beta_{ik}\rangle$ are not known a priori, they are obtained by measurement (e.g.\ Hadamard or swap test).
If the branches are orthonormal, $\mathcal{N}_k$ reduces to a classical norm and no quantum overlap estimation is required.
Qubits measured in $M_c=0$ are drawn as blue lines.
When several classical indices are contracted simultaneously, one first prepares $|a\rangle=\sum_i a_iw_{ik}|i\rangle$ and implements the $U_{a_{ik}}$ as multi-controlled gates before applying a Hadamard gate and measuring the ancilla register in $|0\rangle^{\otimes n}$.

\subsection{Case 3: a classical tensor contracted with a hybrid tensor along a quantum index}

\begin{figure}[htb]
  \centering
  \includegraphics[width=0.5\linewidth]{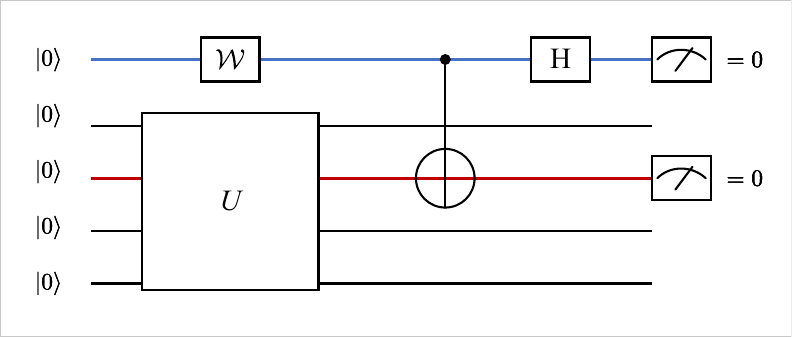}
  \caption{Quantum circuit for contracting a classical tensor with a hybrid tensor along a quantum index (Case~3).
  CNOT and Hadamard gates project the quantum index, implementing the contraction used when absorbing $U$ or $V$ into a quantum connector register.}
  \label{fig:case3}
\end{figure}

Contracting a classical vector $a_i$ with a hybrid tensor $[\mathcal{B}]_k=w_k|\beta_k\rangle$ on its quantum index $q$ gives
\begin{equation}
|\gamma_k\rangle = w_k\sum_i a_i\langle i|\beta_k\rangle.
\end{equation}
Expand $|\beta_k\rangle=b_0|0\rangle_q|\beta_{0k}\rangle+b_1|1\rangle_q|\beta_{1k}\rangle$ on the quantum index $q$, and prepare $a_0|0\rangle_c+a_1|1\rangle_c$.
The circuit in Fig.~\ref{fig:case3} applies $U|0\rangle=|\beta_k\rangle$, then uses CNOT and Hadamard gates to project the quantum index.

The state evolution (key steps only) is
\begin{equation}
\begin{aligned}
|0\rangle_c|0\rangle
\xrightarrow{\mathcal{W}, U}& (a_0|0\rangle_c+a_1|1\rangle_c)\otimes|\beta_k\rangle \\
=&(a_0|0\rangle_c+a_1|1\rangle_c)\otimes(b_0|0\rangle_q|\beta_{0k}\rangle+b_1|1\rangle_q|\beta_{1k}\rangle) \\
\xrightarrow[M_q=0]{\mathrm{CNOT}}& a_0 b_0 |0\rangle_c|\beta_{0k}\rangle + a_1 b_1 |1\rangle_c|\beta_{1k}\rangle \\
\xrightarrow[M_c=0]{H}& a_0 b_0 |\beta_{0k}\rangle + a_1 b_1 |\beta_{1k}\rangle\\
=&a_0 \langle 0|\beta_{k}\rangle + a_1 \langle 1|\beta_{k}\rangle,
\end{aligned}
\end{equation}
where $M_q$ and $M_c$ denote the measurement outcomes on the quantum index $q$ (the qubit represented by the red line) and the ancilla $c$ (the blue line), respectively.
Qubits measured in $M_q=0$ are drawn as red lines.
Post-selection returns a normalized physical state that differs from $|\gamma_k\rangle$ by the norm
\begin{equation}
\mathcal{N}_k=\sqrt{\langle\gamma_k|\gamma_k\rangle}=w_k\sqrt{\sum_{ii'}a_ia_{i'}^*\langle\beta_k|i'\rangle\langle i|\beta_k\rangle}.
\end{equation}
The operators $|i'\rangle\langle i|$ can be expanded in the Pauli basis and estimated by measurement.
For multiple contracted quantum indices, one prepares $|a\rangle=\sum_i a_i|i\rangle$ and applies CNOT gates between the ancilla and the target qubits before the final Hadamard measurement.

\subsection{Case 4: a classical tensor contracted with a hybrid tensor along hybrid indices}

\begin{figure}[htb]
  \centering
  \includegraphics[width=0.6\linewidth]{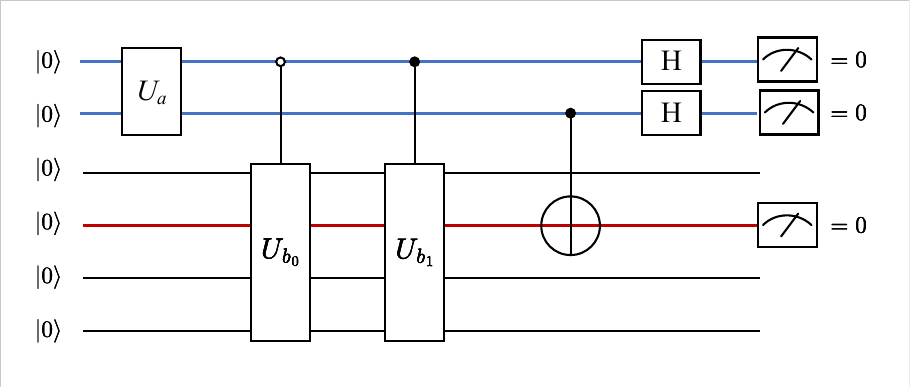}
  \caption{Quantum circuit for contracting a classical tensor with a hybrid tensor along hybrid indices (Case~4).
  An ancilla prepares the weighted classical register $\sum_{ij}a_{ij}w_{ik}|ij\rangle_c$.
  Controlled-$U_b$, CNOT, and Hadamard post-selection implement Eq.~\eqref{eq:case4_gamma}.}
  \label{fig:case4}
\end{figure}

As an example, we contract a rank-$2$ classical tensor $a_{ij}$ with $[\mathcal{B}]_{ik}=w_{ik}|\beta_{ik}\rangle$ on both a classical and a quantum index, giving
\begin{equation}\label{eq:case4_gamma}
|\gamma_k\rangle = \sum_{ij} a_{ij}w_{ik}\langle j|\beta_{ik}\rangle.
\end{equation}
Prepare the classical register state $\sum_{ij}a_{ij}w_{ik}|ij\rangle_c$, and choose unitaries with $U_{b_0}|0\rangle=|\beta_{0k}\rangle$ and $U_{b_1}|0\rangle=|\beta_{1k}\rangle$.
Expanding the hybrid branches on the quantum qubit $q$ corresponding to the contracted quantum index,
\begin{equation}
\begin{aligned}
|\beta_{0k}\rangle &= b_{00}|0\rangle_q|\beta_{00k}\rangle + b_{01}|1\rangle_q|\beta_{01k}\rangle,\\
|\beta_{1k}\rangle &= b_{10}|0\rangle_q|\beta_{10k}\rangle + b_{11}|1\rangle_q|\beta_{11k}\rangle.
\end{aligned}
\end{equation}
The circuit in Fig.~\ref{fig:case4} proceeds as (key steps only)
\begin{equation}
\begin{aligned}
|0\rangle|0\rangle
\xrightarrow{U_a}& \sum_{ij}a_{ij}w_{ik}|ij\rangle_c|0\rangle \\
\xrightarrow{\mathrm{C}\text{-}U_b}&
   \sum_{ij}a_{ij}w_{ik}|ij\rangle_c|\beta_{ik}\rangle\\
=&\sum_{ijl}a_{ij}w_{ik}b_{il}|ij\rangle_c |l\rangle_q|\beta_{ilk}\rangle\\
\xrightarrow[M_q=0]{\mathrm{CNOT}}&
   \sum_{ij}a_{ij}w_{ik}b_{ij}|ij\rangle_c|\beta_{ijk}\rangle\\
\xrightarrow[M_c=0]{H^{\otimes 2}}&
   \sum_{ij}a_{ij}w_{ik}b_{ij}|\beta_{ijk}\rangle\\
=&\sum_{i,j}a_{ij}w_{ik}\langle j|\beta_{ik}\rangle.
\end{aligned}
\end{equation}

Post-selecting the ancilla register in $|00\rangle_c$ returns a normalized physical state that differs from Eq.~\eqref{eq:case4_gamma} by the norm
\begin{equation}
  \mathcal{N}_k=\sqrt{\langle\gamma_k|\gamma_k\rangle}=\sqrt{\sum_{ii'jj'} a_{ij}a_{i'j'}^*w_{ik}w_{i'k}^*\langle \beta_{i'k}|j'\rangle\langle j|\beta_{ik}\rangle}.
\end{equation}
The same pattern extends to larger bond dimensions by using more ancilla qubits and by flattening those dimensions.

\subsection{Case 5: a hybrid tensor contracted with a hybrid tensor along classical indices}

\begin{figure}[htb]
  \centering
  \includegraphics[width=0.5\linewidth]{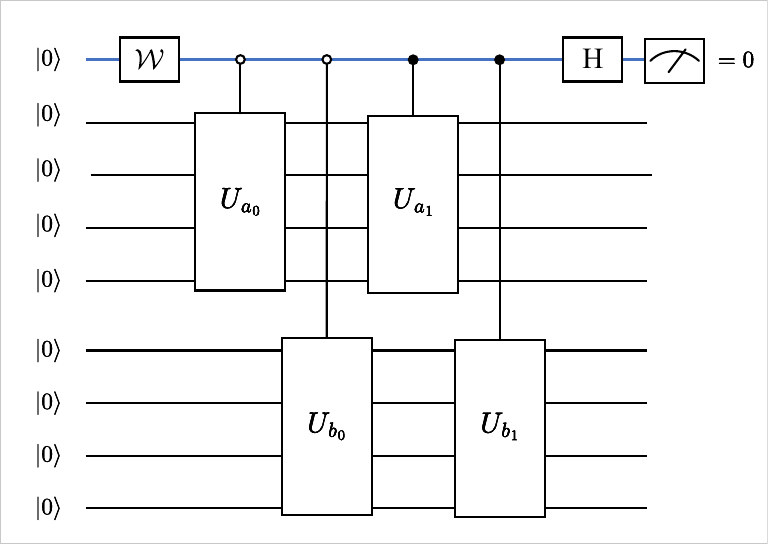}
  \caption{Quantum circuit for contracting two hybrid tensors along classical indices (Case~5).
  An ancilla prepared by $\mathcal{W}$ indexes the classical branch pair $(|\alpha_{ij}\rangle,|\beta_{jk}\rangle)$.
  Measuring $M_c=0$ recovers Eq.~\eqref{eq:case5_gamma}.}
  \label{fig:case5}
\end{figure}

Contracting $[\mathcal{A}]_{ij}=w_{ij}|\alpha_{ij}\rangle$ and $[\mathcal{B}]_{jk}=v_{jk}|\beta_{jk}\rangle$ along the shared classical index $j$, for fixed boundary indices $i$ and $k$, gives
\begin{equation}\label{eq:case5_gamma}
|\gamma_{ik}\rangle = \sum_j w_{ij}v_{jk}|\alpha_{ij}\rangle|\beta_{jk}\rangle.
\end{equation}
The unitary $\mathcal{W}$ prepares $\sum_{j}w_{ij}v_{jk}|j\rangle_c$.
Choose unitaries satisfying $U_{a_0}|0\rangle=|\alpha_{i0}\rangle$, $U_{a_1}|0\rangle=|\alpha_{i1}\rangle$, $U_{b_0}|0\rangle=|\beta_{0k}\rangle$, and $U_{b_1}|0\rangle=|\beta_{1k}\rangle$.
The circuit in Fig.~\ref{fig:case5} uses an ancilla to select the classical branch.

The state evolution (key steps only) is
\begin{equation}
\begin{aligned}
|0\rangle_c|0\rangle|0\rangle
\xrightarrow{\mathcal{W}}& \sum_j w_{ij}v_{jk}|j\rangle_c|0\rangle|0\rangle \\
\xrightarrow{\mathrm{C}\text{-}U}& \sum_j w_{ij}v_{jk}|j\rangle_c|\alpha_{ij}\rangle|\beta_{jk}\rangle\\
\xrightarrow[M_c=0]{H}& \sum_j w_{ij}v_{jk}|\alpha_{ij}\rangle|\beta_{jk}\rangle.
\end{aligned}
\end{equation}
where $M_c=0$ denotes post-selection of the ancilla in $|0\rangle$.
Post-selection returns a normalized physical state that differs from Eq.~\eqref{eq:case5_gamma} by the norm
\begin{equation}
  \mathcal{N}_{ik}=\sqrt{\langle\gamma_{ik}|\gamma_{ik}\rangle}=\sqrt{\sum_{jj'} w_{ij}w_{ij'}^*v_{jk}v_{j'k}^*\langle\alpha_{ij'}|\alpha_{ij}\rangle\langle\beta_{j'k}|\beta_{jk}\rangle}.
\end{equation}
For several classical indices contracted at once, enlarge the ancilla register, prepare the joint weighted superposition with $\mathcal{W}$, and use multi-controlled unitaries before measuring $M_c=0$ on the ancilla register.

\subsection{Case 6: a hybrid tensor contracted with a hybrid tensor along a classical and a quantum index}

\begin{figure}[htb]
  \centering
  \includegraphics[width=0.5\linewidth]{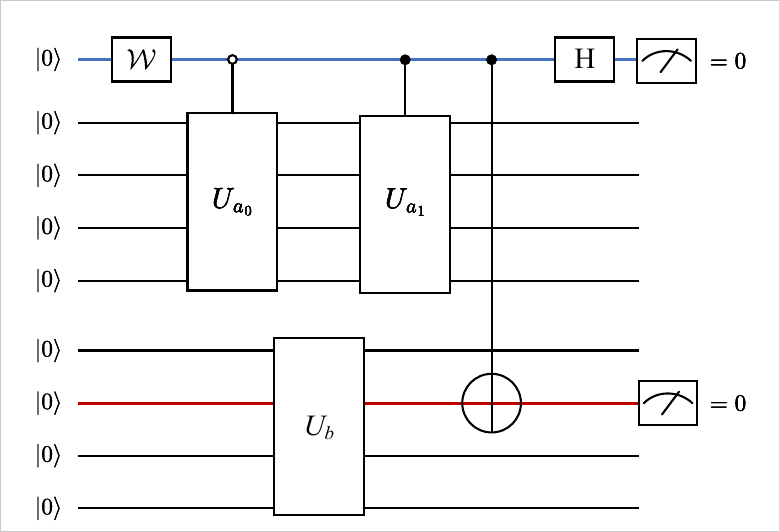}
  \caption{Quantum circuit for contracting hybrid tensors along a classical index and a quantum index (Case~6).
  Controlled unitaries select the classical branch.
  Post-selection with $M_q=0$ after CNOT and $M_c=0$ after Hadamard projects onto Eq.~\eqref{eq:case6_gamma}.}
  \label{fig:case6}
\end{figure}

Contracting $[\mathcal{A}]_{ij}=w_{ij}|\alpha_{ij}\rangle$ and $[\mathcal{B}]_{k}=v_{k}|\beta_{k}\rangle$ along classical index $j$ and quantum index corresponding to qubit $q$, for fixed indices $i$ and $k$, yields
\begin{equation}\label{eq:case6_gamma}
|\gamma_{ik}\rangle = v_k\sum_j w_{ij}|\alpha_{ij}\rangle\langle j|\beta_k\rangle,
\end{equation}
with $U_{a_0}|0\rangle=|\alpha_{i0}\rangle$, $U_{a_1}|0\rangle=|\alpha_{i1}\rangle$, and $U_b|0\rangle=|\beta_k\rangle$.
Expand
\begin{equation}
|\beta_k\rangle=b_0|0\rangle_q|\beta_{0k}\rangle+b_1|1\rangle_q|\beta_{1k}\rangle.
\end{equation}
Gate $\mathcal{W}$ prepares $w_{i0}v_k|0\rangle+w_{i1}v_k|1\rangle$.
The circuit is shown in Fig.~\ref{fig:case6}.

The key steps of the state evolution are
\begin{equation}
\begin{aligned}
|0\rangle_c|0\rangle|0\rangle
\xrightarrow{\mathcal{W}}& \sum_j w_{ij}v_k|j\rangle_c|0\rangle|0\rangle \\
\xrightarrow{\mathrm{C}\text{-}U_a,\,U_b}&v_k\sum_{j}w_{ij}|j\rangle_c|\alpha_{ij}\rangle|\beta_k\rangle\\
=& v_k\sum_{jl}w_{ij}|j\rangle_c|\alpha_{ij}\rangle b_l|l\rangle_q|\beta_{lk}\rangle\\
\xrightarrow[M_q=0]{\mathrm{CNOT}}&v_k\sum_j w_{ij}b_j|j\rangle_c|\alpha_{ij}\rangle|\beta_{jk}\rangle\\
\xrightarrow[M_c=0]{H}&v_k\sum_j w_{ij}b_j|\alpha_{ij}\rangle|\beta_{jk}\rangle\\
=& v_k\sum_j w_{ij}|\alpha_{ij}\rangle\langle j|\beta_k\rangle,
\end{aligned}
\end{equation}
where $M_q=0$ and $M_c=0$ are the projections on the quantum index and the ancilla, respectively.
Post-selection returns a normalized physical state that differs from Eq.~\eqref{eq:case6_gamma} by the norm
\begin{equation}
  \mathcal{N}_{ik}=\sqrt{\langle\gamma_{ik}|\gamma_{ik}\rangle}=v_k\sqrt{
    \sum_{jj'} w_{ij}w_{ij'}^*\langle\alpha_{ij'}|\alpha_{ij}\rangle\langle\beta_k|j'\rangle\langle j|\beta_k\rangle
  }.
\end{equation}
Multiple contracted indices follow by enlarging the control register and the CNOT fan-out before the same measurements.

\subsection{Case 7: a hybrid tensor contracted with a hybrid tensor along classical and hybrid indices}

\begin{figure}[htb]
  \centering
  \includegraphics[width=0.75\linewidth]{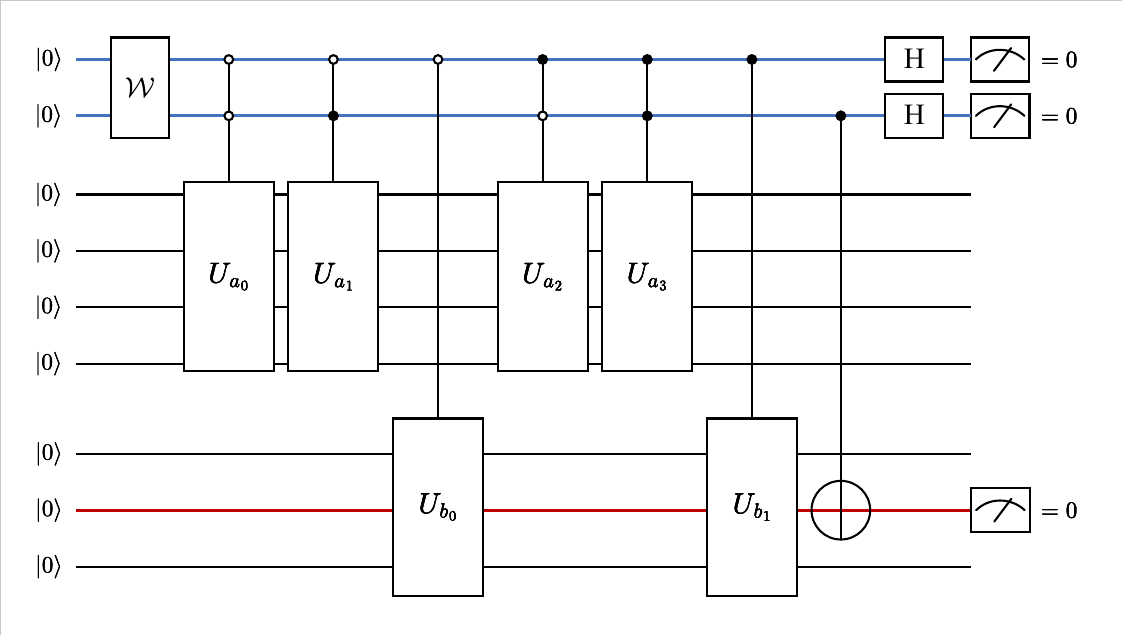}
  \caption{Quantum circuit for contracting hybrid tensors along classical indices and hybrid indices (Case~7).
  Two ancilla qubits index $(i,j)$.
  Post-selection with $M_q=0$ and $M_c=0$ implements Eq.~\eqref{eq:case7_gamma}.}
  \label{fig:case7}
\end{figure}

For $[\mathcal{A}]_{ij}=w_{ij}|\alpha_{ij}\rangle$ and $[\mathcal{B}]_{ik}=v_{ik}|\beta_{ik}\rangle$ contracted along the classical indices $i,j$ of $[\mathcal{A}]$ and the classical index $i$ with quantum index corresponding to qubit $q$ of $[\mathcal{B}]$, fixing index $k$,
\begin{equation}\label{eq:case7_gamma}
|\gamma_k\rangle = \sum_{ij} w_{ij}v_{ik}|\alpha_{ij}\rangle\langle j|\beta_{ik}\rangle.
\end{equation}
Label the four classical pairs by a two-qubit ancilla and take $U_{a_{ij}}|0\rangle=|\alpha_{ij}\rangle$ together with
\begin{equation}
\begin{aligned}
U_{b_0}|0\rangle=|\beta_{0k}\rangle&=b_{00}|0\rangle_q|\beta_{00k}\rangle+b_{01}|1\rangle_q|\beta_{01k}\rangle,\\
U_{b_1}|0\rangle=|\beta_{1k}\rangle&=b_{10}|0\rangle_q|\beta_{10k}\rangle+b_{11}|1\rangle_q|\beta_{11k}\rangle.
\end{aligned}
\end{equation}
The two-qubit gate $\mathcal{W}$ prepares $\sum_{ij} w_{ij}v_{ik}|ij\rangle_c$.
The circuit is shown in Fig.~\ref{fig:case7}.

The key step of the evolution is
\begin{equation}
\begin{aligned}
|0\rangle_c|0\rangle|0\rangle
\xrightarrow{\mathcal{W},\mathrm{C}\text{-}U}&
  \sum_{ij}w_{ij}v_{ik}|ij\rangle_c|\alpha_{ij}\rangle|\beta_{ik}\rangle \\
=&\sum_{ijl} w_{ij}v_{ik}|ij\rangle_c|\alpha_{ij}\rangle b_{il}|l\rangle_q|\beta_{ilk}\rangle\\
\xrightarrow[M_q=0]{\mathrm{CNOT}}&
  \sum_{ij}w_{ij}v_{ik}b_{ij}|ij\rangle_c|\alpha_{ij}\rangle|\beta_{ijk}\rangle \\
\xrightarrow[M_c=0]{H^{\otimes 2}}&
  \sum_{ij}w_{ij}v_{ik}b_{ij}|\alpha_{ij}\rangle|\beta_{ijk}\rangle\\
=&\sum_{ij}w_{ij}v_{ik}|\alpha_{ij}\rangle\langle j|\beta_{ik}\rangle.
\end{aligned}
\end{equation}
Measuring $M_q=0$ and $M_c=0$ recovers the contracted branch.
Post-selection returns a normalized physical state that differs from Eq.~\eqref{eq:case7_gamma} by the norm
\begin{equation}
  \mathcal{N}_k=\sqrt{\langle\gamma_k|\gamma_k\rangle}=\sqrt{
    \sum_{ii'jj'} w_{ij}w_{i'j'}^*v_{ik}v_{i'k}^*\langle\alpha_{i'j'}|\alpha_{ij}\rangle\langle\beta_{i'k}|j'\rangle\langle j|\beta_{ik}\rangle
  }.
\end{equation}

\subsection{Case 8: a hybrid tensor contracted with a hybrid tensor along quantum indices}

\begin{figure}[htb]
  \centering
  \includegraphics[width=0.5\linewidth]{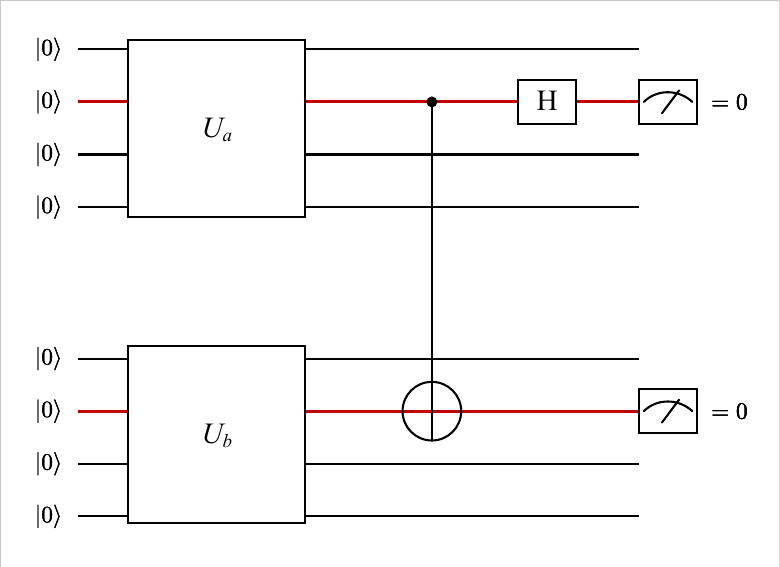}
  \caption{Quantum circuit for contracting two hybrid tensors along quantum indices (Case~8).
  CNOT and Hadamard gates with $M_q=0$ give the result in Eq.~\eqref{eq:case8_gamma}.}
  \label{fig:case8}
\end{figure}

Contracting two hybrid tensors $\mathcal{A}=w|\alpha\rangle$ and $\mathcal{B}=v|\beta\rangle$ along matched quantum indices gives the residual state
\begin{equation}\label{eq:case8_gamma}
|\gamma\rangle = wv\sum_i \langle i|\alpha\rangle\langle i|\beta\rangle,
\end{equation}
where $\langle i|\alpha\rangle$ and $\langle i|\beta\rangle$ are understood as the states of the remaining qubits.
Expand
\begin{equation}
|\alpha\rangle=a_0|0\rangle_{q_a}|\alpha_0\rangle+a_1|1\rangle_{q_a}|\alpha_1\rangle,\qquad
|\beta\rangle=b_0|0\rangle_{q_b}|\beta_0\rangle+b_1|1\rangle_{q_b}|\beta_1\rangle.
\end{equation}
The circuit is shown in Fig.~\ref{fig:case8}.

It prepares both states and projects the shared quantum indices (key steps only):
\begin{equation}
\begin{aligned}
|0\rangle|0\rangle
\xrightarrow{U_a,\,U_b}& |\alpha\rangle|\beta\rangle \\
=&\sum_{ij} a_ib_j|i\rangle_{q_a}|j\rangle_{q_b}|\alpha_i\rangle|\beta_j\rangle\\
\xrightarrow[M_q=0]{\mathrm{CNOT}, H}& \sum_i a_ib_i|\alpha_i\rangle|\beta_i\rangle\\
=& \sum_i\langle i|\alpha\rangle\langle i|\beta\rangle.
\end{aligned}
\end{equation}
Here both matched quantum indices are treated as the measured quantum register ($M_q$).
Post-selection returns a normalized physical state that differs from Eq.~\eqref{eq:case8_gamma} by the norm
\begin{equation}\label{eq:norm_case_8}
  \mathcal{N}=\sqrt{\langle\gamma|\gamma\rangle}=wv\sqrt{\sum_{ij} \langle\alpha|i\rangle\langle j|\alpha\rangle\cdot\langle\beta|i\rangle\langle j|\beta\rangle}.
\end{equation}
The terms in the form of $|i\rangle\langle j|$ can be written as linear combinations of Pauli operators, so this norm can be measured.
For several quantum indices, apply CNOT between each matched pair and measure the corresponding registers with $M_q=0$ and Hadamard gates.

\subsection{Case 9: a hybrid tensor contracted with a hybrid tensor along quantum and hybrid indices}

\begin{figure}[htb]
  \centering
  \includegraphics[width=0.75\linewidth]{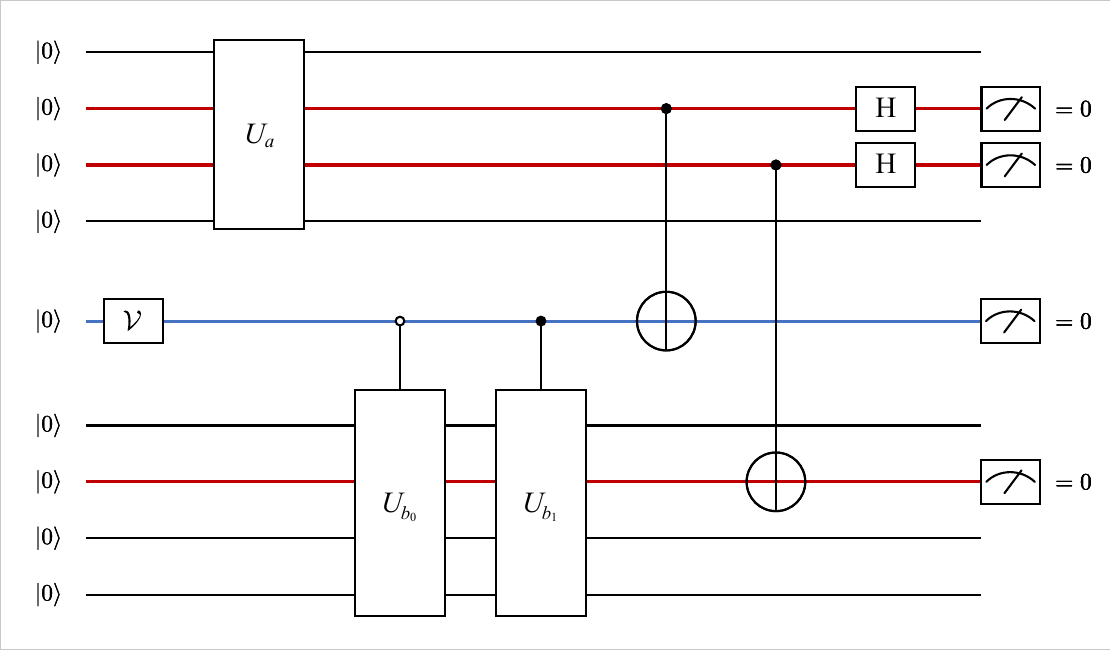}
  \caption{Quantum circuit for contracting hybrid tensors along quantum indices and hybrid indices (Case~9).
  After the CNOT layer one measures $M_c=0$ to drop the classical-branch ancilla, then $M_q=0$ after Hadamard on the quantum-index register, implementing Eq.~\eqref{eq:case9_gamma}.}
  \label{fig:case9}
\end{figure}

Contracting $\mathcal{A}=w|\alpha\rangle$ with $[\mathcal{B}]_{ik}=v_{ik}|\beta_{ik}\rangle$ along quantum indices (with corresponding qubits labeled as $q_a$) of $\mathcal{A}$ and classical index $i$ and another quantum index (qubit labeled as $q_b$) of $\mathcal{B}$ gives
\begin{equation}\label{eq:case9_gamma}
|\gamma_k\rangle = w\sum_{ij} v_{ik}\langle ij|\alpha\rangle\langle j|\beta_{ik}\rangle.
\end{equation}
Prepare $U_a|0\rangle=|\alpha\rangle$, which can be expanded along indices $i$ and $j$.
\begin{equation}
  |\alpha\rangle=
    a_{00}|00\rangle_{q_a}|\alpha_{00}\rangle+a_{01}|01\rangle_{q_a}|\alpha_{01}\rangle
    +a_{10}|10\rangle_{q_a}|\alpha_{10}\rangle+a_{11}|11\rangle_{q_a}|\alpha_{11}\rangle,
  \end{equation}
and write $|\beta_{ik}\rangle$ as
\begin{equation}
  \begin{aligned}
    |\beta_{0k}\rangle&=b_{00}|0\rangle_{q_b}|\beta_{00k}\rangle+b_{01}|1\rangle_{q_b}|\beta_{01k}\rangle,\\
    |\beta_{1k}\rangle&=b_{10}|0\rangle_{q_b}|\beta_{10k}\rangle+b_{11}|1\rangle_{q_b}|\beta_{11k}\rangle.
  \end{aligned}
\end{equation}
The circuit is shown in Fig.~\ref{fig:case9}.

It applies $U_a$, an ancilla prepared in $v_{0k}|0\rangle + v_{1k}|1\rangle$, a controlled pair $\{U_{b_0},U_{b_1}\}$, and CNOT gates on the matched registers.
Measuring the classical-branch ancilla first removes that register and retains only matching labels $i$.
The subsequent Hadamard measurement then projects the quantum-index register $|ij\rangle$.
The state evolution (key steps only) is
\begin{equation}
\begin{aligned}
|0\rangle|0\rangle_c|0\rangle
\xrightarrow{U_a,\,\mathcal{V},\,\mathrm{C}\text{-}U_b}&
  |\alpha\rangle\otimes\bigl(v_{0k}|0\rangle_c|\beta_{0k}\rangle+v_{1k}|1\rangle_c|\beta_{1k}\rangle\bigr) \\
=& \sum_{ij}a_{ij}|ij\rangle_{q_a}|\alpha_{ij}\rangle \otimes \sum_lv_{lk}|l\rangle_c|\beta_{lk}\rangle\\
\xrightarrow[M_c=0]{\mathrm{CNOT}_c}&\sum_{ij}a_{ij}v_{ik}|ij\rangle_{q_a}|\alpha_{ij}\rangle|\beta_{ik}\rangle\\
&=\sum_{ijl} a_{ij}v_{ik}|ij\rangle_{q_a}|\alpha_{ij}\rangle b_{il}|l\rangle_{q_b}|\beta_{ilk}\rangle\\
\xrightarrow[M_{q_b}=0]{\mathrm{CNOT}_q}&\sum_{ij} a_{ij}v_{ik}|ij\rangle_{q_a}|\alpha_{ij}\rangle b_{ij}|\beta_{ijk}\rangle\\
\xrightarrow[M_{q_a}=0]{H^{\otimes 2}}&
  \sum_{ij}a_{ij}b_{ij}v_{ik}|\alpha_{ij}\rangle|\beta_{ijk}\rangle \\
=& \sum_{ij}v_{ik}\langle ij|\alpha\rangle\langle j|\beta_{ik}\rangle.
\end{aligned}
\end{equation}
Post-selection returns a normalized physical state that differs from Eq.~\eqref{eq:case9_gamma} by the norm
\begin{equation}
  \mathcal{N}_k=\sqrt{\langle\gamma_k|\gamma_k\rangle}=w\sqrt{
    \sum_{ii'jj'} v_{ik}v_{i'k}^*\langle\alpha|i'j'\rangle\langle ij|\alpha\rangle\langle\beta_{i'k}|j'\rangle\langle j|\beta_{ik}\rangle
  }.
\end{equation}
Additional contracted indices enlarge the CNOT and Hadamard registers in the same way.

\subsection{Case 10: a hybrid tensor contracted with a hybrid tensor along hybrid indices}

When both contracted indices are hybrid, two index orderings arise.

\begin{figure}[t]
  \centering
  \includegraphics[width=0.75\linewidth]{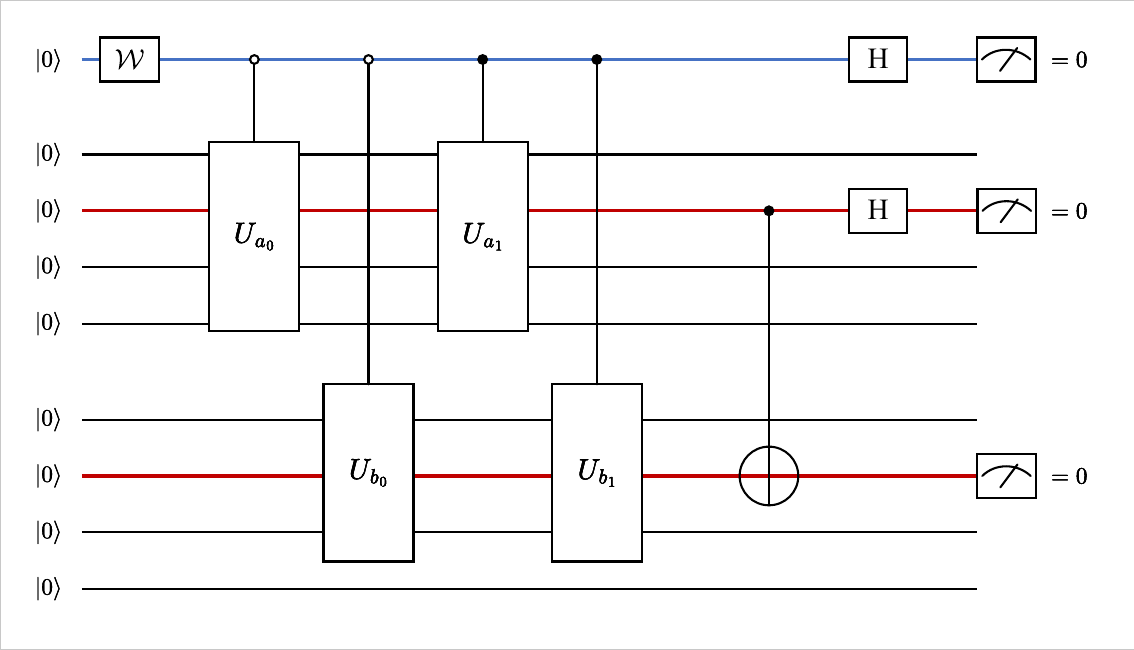}
  \caption{Quantum circuit for contracting two hybrid tensors along hybrid indices with classical--classical and quantum--quantum pairing (Case~10a).
  Measurements $M_q=0$ and $M_c=0$ implement Eq.~\eqref{eq:case10a_gamma}.}
  \label{fig:case10_1}
\end{figure}

\paragraph{Classical--classical and quantum--quantum pairing.} Contracting $[\mathcal{A}]_j=w_j|\alpha_j\rangle$ and $[\mathcal{B}]_j=v_j|\beta_j\rangle$ along matching classical and quantum indices (with corresponding qubits labeled as $q_a$ and $q_b$) gives
\begin{equation}\label{eq:case10a_gamma}
|\gamma\rangle = \sum_{ij} w_jv_j\langle i|\alpha_j\rangle\langle i|\beta_j\rangle.
\end{equation}
The unitary $\mathcal{W}$ prepares $w_0v_0|0\rangle+w_1v_1|1\rangle$.
With $U_{a_j}|0\rangle=|\alpha_j\rangle=a_{j0}|0\rangle_{q_a}|\alpha_{j0}\rangle+a_{j1}|1\rangle_{q_a}|\alpha_{j1}\rangle$ and likewise for $U_{b_j}|0\rangle=|\beta_j\rangle=b_{j0}|0\rangle_{q_b}|\beta_{j0}\rangle+b_{j1}|1\rangle_{q_b}|\beta_{j1}\rangle$, the circuit in Fig.~\ref{fig:case10_1} evolves as (key steps only)
\begin{equation}
\begin{aligned}
|0\rangle_c|0\rangle|0\rangle
\xrightarrow{\mathcal{W}}&\bigl(w_0v_0|0\rangle_c+w_1v_1|1\rangle_c\bigr)|0\rangle|0\rangle \\
\xrightarrow{\mathrm{C}\text{-}U}&\sum_{j}w_jv_j|j\rangle_c|\alpha_j\rangle|\beta_j\rangle\\
=&\sum_{ijk}w_jv_j|j\rangle_c a_{ji}|i\rangle_{q_a}|\alpha_{ji}\rangle b_{jk}|k\rangle_{q_b}|\beta_{jk}\rangle\\
\xrightarrow[M_{q_b}=0]{\mathrm{CNOT}}&\sum_{ij}w_jv_j|j\rangle_c |i\rangle_{q_a} a_{ji}b_{ji}|\alpha_{ji}\rangle|\beta_{ji}\rangle\\
\xrightarrow[M_{c,q_a}=0]{H}& \sum_{ij}w_jv_ja_{ji}b_{ji}|\alpha_{ji}\rangle|\beta_{ji}\rangle\\
=& \sum_{ij}w_jv_j\langle i|\alpha_j\rangle\langle i|\beta_j\rangle.
\end{aligned}
\end{equation}
Here $M_q=0$ (including $M_{q_a}$ and $M_{q_b}$) projects the matched quantum indices and $M_c=0$ post-selects the classical-branch ancilla.
Post-selection returns a normalized physical state that differs from Eq.~\eqref{eq:case10a_gamma} by the norm
\begin{equation}
  \mathcal{N}=\sqrt{\langle\gamma|\gamma\rangle}=\sqrt{
    \sum_{ii'jj'} w_jw_{j'}^*v_jv_{j'}^*\langle \alpha_{j'}|i'\rangle\langle i|\alpha_j\rangle\langle\beta_{j'}|i'\rangle\langle i|\beta_j\rangle
  }.
\end{equation}

\begin{figure}[t]
  \centering
  \includegraphics[width=0.75\linewidth]{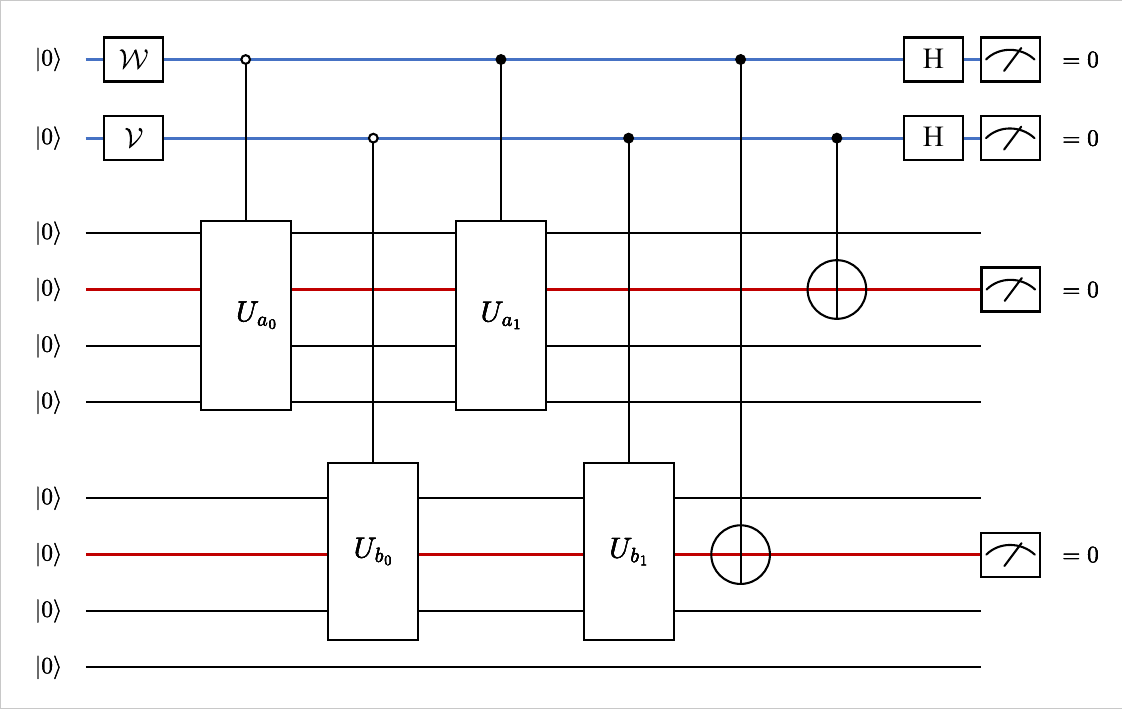}
  \caption{Quantum circuit for contracting two hybrid tensors along hybrid indices with classical--quantum and quantum--classical pairing (Case~10b).
  Crossed CNOT routing with $M_q=0$ and $M_c=0$ implements Eq.~\eqref{eq:case10b_gamma}.}
  \label{fig:case10_2}
\end{figure}

\paragraph{Classical--quantum and quantum--classical pairing.} When the classical index of one tensor pairs with the quantum index of the other, the contraction reads
\begin{equation}\label{eq:case10b_gamma}
|\gamma\rangle = \sum_{ij} w_jv_i\langle i|\alpha_j\rangle\langle j|\beta_i\rangle.
\end{equation}
The unitaries $\mathcal{W}$ and $\mathcal{V}$ prepare $w_0|0\rangle+w_1|1\rangle$ and $v_0|0\rangle+v_1|1\rangle$, respectively.
Two ancillas select the crossed classical labels.
With the same branch expansions as above, the circuit in Fig.~\ref{fig:case10_2} evolves as (key steps only)
\begin{equation}
\begin{aligned}
|0\rangle_{c_1}|0\rangle_{c_2}|0\rangle|0\rangle
\xrightarrow{\mathcal{W}, \mathcal{V}}&
  \sum_{ij}w_jv_i|j\rangle_{c_1}|i\rangle_{c_2}|0\rangle|0\rangle \\
\xrightarrow{\mathrm{C}\text{-}U}&
  \sum_{ij}w_jv_i|j\rangle_{c_1}|i\rangle_{c_2}|\alpha_j\rangle|\beta_i\rangle\\
&=\sum_{ijkl}w_jv_i|j\rangle_{c_1}|i\rangle_{c_2}a_{jk}|k\rangle_{q_a}|\alpha_{jk}\rangle b_{il}|l\rangle_{q_b}|\beta_{il}\rangle\\
\xrightarrow[M_q=0]{\mathrm{CNOT}}&\sum_{ij}w_jv_ia_{ji}b_{ij}|j\rangle_{c_1}|i\rangle_{c_2}|\alpha_{ji}\rangle|\beta_{ij}\rangle\\
\xrightarrow[M_c=0]{H^{\otimes 2}}&\sum_{ij}w_jv_ia_{ji}b_{ij}|\alpha_{ji}\rangle|\beta_{ij}\rangle\\
=&\sum_{ij}w_jv_i\langle i|\alpha_j\rangle \langle j|\beta_i\rangle.
\end{aligned}
\end{equation}
Measuring $M_q=0$ on the crossed quantum indices and $M_c=0$ on the two ancillas implements the crossed coefficient pattern that distinguishes this pairing from Case~10a.
Post-selection returns a normalized physical state that differs from Eq.~\eqref{eq:case10b_gamma} by the norm
\begin{equation}
  \mathcal{N}=\sqrt{\langle\gamma|\gamma\rangle}=\sqrt{
    \sum_{ii'jj'} w_jw_{j'}^*v_iv_{i'}^*\langle\alpha_{j'}|i'\rangle\langle i|\alpha_j\rangle\langle\beta_{i'}|j'\rangle\langle j|\beta_i\rangle
  }.
\end{equation}

In all cases above, the weights of the contracted tensors enter the circuit as preparation parameters where needed.
Post-selection leaves a normalized physical state.
The norm $\mathcal{N}=\sqrt{\langle\gamma|\gamma\rangle}$ matches the unnormalized written contraction $|\gamma\rangle$.
In general $\mathcal{N}$ is obtained by estimating the overlaps or Pauli terms that appear in $\langle\gamma|\gamma\rangle$.
Those estimates require quantum measurements whenever the physical states on the contracted registers enter the expression.
This step can be costly.
In the main-text Trotter protocol, Theorems~1 and~2 keep branch norms at unity after each orthogonalization and connector update, so explicit weight recording may be omitted.

\section{Using quantum indices for connection}\label{sec:quantum_connection}

In the default main-text layout, each hybrid site carries classical connector indices $\mathbf{c}$ in addition to on-site qubits $\mathbf{q}$.
Here we describe an equivalent layout in which connector labels on each hybrid site are represented by quantum registers.
Every branch state can then remain in a single entangled register.
The classical connecting tensor $\boldsymbol{\lambda}$ continues to store the decomposition coefficients $\lambda_i$ and, after truncation, the Schmidt coefficients $s_l$.
Because $\boldsymbol{\lambda}$ remains classical, the orthogonalization and SVD steps of the main text apply without modification.
Theorems~1 and~2 hold unchanged once the quantum connector register $|i\rangle$ is identified with the classical label $i$.

A site with quantum connectors and physical qubits $q_{\rm phys}$ is written $[\mathcal{A}]^{\mathbf{q}}$, where $\mathbf{q}=(q_1,\ldots,q_{k},q_{\rm phys})$ and $k$ is the number of connection registers on that site (the site degree).
Each connection register $q_j$ uses
\begin{equation}
  n_\chi \equiv \lceil \log_2 \chi \rceil
\end{equation}
qubits, encoding bond labels $i\in\{0,\ldots,\chi-1\}$ in the computational basis $|i\rangle$.
A remote gate of decomposition rank $\rg$ introduces $\lceil\log_2 \rg \rceil$ ancilla qubits per affected site to label the $\rg$ terms in Eq.~\eqref{eq:decomp} of the main text.
These can be merged with existing connection registers when the sites were already connected.

\subsection{Applying remote gates}

A remote two-body unitary admits the same decomposition as in the main text,
\begin{equation}
  U=\sum_{m=1}^{\rg} \lambda_m \,[U_A]_m\otimes [U_B]_m.
\end{equation}
In the quantum-index layout the branch label $m$ is stored on an ancilla register rather than as a classical index $\mathbf{c}$.
One introduces $\lceil\log_2 \rg\rceil$ ancilla qubits per affected site.
Without loss of generality, we assume that $\rg$ is a power of two.
Hadamard gates place that register in a uniform superposition over computational-basis labels $|m\rangle$.
Controlled gates $\mathrm{C}_m\text{-}[U_A]_m$ (respectively $\mathrm{C}_m\text{-}[U_B]_m$) then enact $[U_A]_m$ on the physical qubits when the ancilla reads $|m\rangle$.
Writing $\nrg=\lceil\log_2 \rg\rceil$, the circuit is unitary and produces
\begin{equation}\label{eq:quantum_remote}
\begin{aligned}
  \relax[\mathcal{A}']^{q_{\rm conn},\,q_{\rm phys}}
  &= \Biggl(\prod_{m=1}^{\rg}\mathrm{C}_m\text{-}[U_A]_m\Biggr)H^{\otimes \nrg}|0\rangle^{\otimes \nrg}|\psi\rangle
  = 2^{-\nrg/2}\sum_{m=1}^{\rg} |m\rangle_{q_{\rm conn}}\,[U_A]_m|\psi\rangle,\\
  \relax[\mathcal{B}']^{q_{\rm conn},\,q_{\rm phys}}
  &= \Biggl(\prod_{m=1}^{\rg}\mathrm{C}_m\text{-}[U_B]_m\Biggr)H^{\otimes \nrg}|0\rangle^{\otimes \nrg}|\phi\rangle
  = 2^{-\nrg/2}\sum_{m=1}^{\rg} |m\rangle_{q_{\rm conn}}\,[U_B]_m|\phi\rangle.
\end{aligned}
\end{equation}
The decomposition coefficients $\lambda_m$ are not placed in these kets.
They remain on the classical connecting tensor $\boldsymbol{\lambda}=\mathrm{Diag}(\lambda_1,\ldots,\lambda_{\rg})$.

Hardware stores the normalized registers in Eq.~\eqref{eq:quantum_remote}.
The missing scale is restored by the weight of each hybrid site,
\begin{equation}\label{eq:quantum_remote_weight}
  w\leftarrow 2^{\nrg/2}\,w.
\end{equation}

As a concrete example, consider a remote gate for a two-body $XX$ term
\begin{equation}
  U=\alpha_0\,I\otimes I+\alpha_1\,X\otimes X,
\end{equation}
for which $\rg=2$ and a single ancilla suffices ($\nrg=1$).
If the sites are initially unconnected, $\mathcal{A}=|\psi\rangle$ and $\mathcal{B}=|\phi\rangle$, the circuit consists of a Hadamard on the ancilla followed by $\mathrm{C}_1\text{-}X$:
\begin{equation}
\begin{aligned}
  \relax[\mathcal{A}']
  &= (\mathrm{C}_1\text{-}X)\cdot H\,|0\rangle|\psi\rangle
  = 2^{-1/2}\bigl(|0\rangle|\psi\rangle+|1\rangle X|\psi\rangle\bigr),\\
  \relax[\mathcal{B}']
  &= (\mathrm{C}_1\text{-}X)\cdot H\,|0\rangle|\phi\rangle
  = 2^{-1/2}\bigl(|0\rangle|\phi\rangle+|1\rangle X|\phi\rangle\bigr),
\end{aligned}
\end{equation}
while $\boldsymbol{\lambda}=\mathrm{Diag}(\alpha_0,\alpha_1)$ stores the decomposition coefficients.
The unnormalized targets are $|\gamma_A\rangle=|0\rangle|\psi\rangle+|1\rangle X|\psi\rangle$ and likewise for $B$, so each site updates $w$ by $\sqrt{2}\,w$ after the gate.
Both sites thereby acquire one ancilla as a connection register.
If the sites were already connected, this ancilla is appended to the existing connection registers and $\boldsymbol{\lambda}$ is merged with the prior connector as $\boldsymbol{\lambda}\otimes\boldsymbol{\lambda}'$ before truncation.

\subsection{Truncation}

Truncation follows the main-text procedure, with the cut index $i$ running over the computational basis of the relevant connection register $q_j$.

For a hybrid site $[\mathcal{A}]^{\mathbf{q}}$, orthonormalization with respect to the connection register $q_j$ begins with the Gram matrix
\begin{equation}\label{eq:gram_quantum}
\begin{aligned}
  G_{ii'} &= \sum_{\mathbf{q}^{-}, q_{\rm phys}}
  [\mathcal{A}^*]^{\mathbf{q}^{-}, i,\, q_{\rm phys}}
  [\mathcal{A}]^{\mathbf{q}^{-}, i',\, q_{\rm phys}} \\
  &= \sum_{\mathbf{q}^{-}} \langle \psi_{\mathbf{q}^{-}, i}| \psi_{\mathbf{q}^{-}, i'} \rangle,
\end{aligned}
\end{equation}
where $\mathbf{q}^{-}$ denotes the remaining connection registers other than $q_j$, and $q_{\rm phys}$ labels the physical qubits of the site.
This is Eq.~\eqref{eq:gram_matrix} of the main text with $\mathbf{c}^{-},i$ replaced by $\mathbf{q}^{-},i$.
The matrix $G$ is measured on hardware by expanding $|i\rangle\langle i'|$ in the Pauli basis and estimating the corresponding expectations.
Theorem~1 then defines the mapping tensor $\mathcal{R}$, and the orthogonalized site $[\mathcal{A}_\perp]^{\mathbf{q}}$ is obtained as in the main text.

\begin{figure}[htb]
  \centering
  \includegraphics[width=0.45\linewidth]{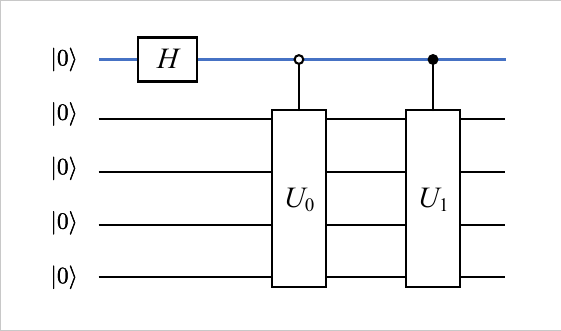}
  \caption{Quantum circuit for converting a classical index to a quantum index.
  Each controlled unitary $U_i$ loads the branch state $|\psi_i\rangle$ onto the physical qubits, leaving the label $i$ on the connection register.}
  \label{fig:c2q}
\end{figure}

Given $[\mathcal{A}_\perp]^{\mathbf{q}_A}$ and $[\mathcal{B}_\perp]^{\mathbf{q}_B}$ linked by the classical connector $\boldsymbol{\lambda}$, the update to $\boldsymbol{\lambda}_\perp$ and the SVD with $(\boldsymbol{\lambda}_\perp)_{kk'}=\sum_l U_{kl}s_l V_{lk'}$ are identical to the main text.
Absorbing $U$ and $V$ uses the combined classical matrices $\mathcal{R}_AU$ and $\mathcal{R}_BV^T$ along the quantum connection register.
Each contraction is implemented by the Case~3 circuit in Sec.~I (a classical tensor contracted with a hybrid tensor along a quantum index):
\begin{equation}\label{eq:contraction_quantum_index}
  \begin{aligned}
    \relax[\widetilde{\mathcal{A}}]^{\mathbf{q}^-_A,\, q_{\rm phys}}_{l} &= \sum_{i,j} [\mathcal{A}]^{\mathbf{q}^-_A,\, i,\, q_{\rm phys}}(\mathcal{R}_A)_{ij} U_{jl} = \sum_{i} [\mathcal{A}]^{\mathbf{q}^-_A,\, i,\, q_{\rm phys}}(\mathcal{R}_A U)_{il}, \\
    \relax[\widetilde{\mathcal{B}}]^{\mathbf{q}^-_B,\, q_{\rm phys}}_{l} &= \sum_{i,j} [\mathcal{B}]^{\mathbf{q}^-_B,\, i,\, q_{\rm phys}}(\mathcal{R}_B)_{ij} V_{lj} = \sum_{i} [\mathcal{B}]^{\mathbf{q}^-_B,\, i,\, q_{\rm phys}}(\mathcal{R}_B V^T)_{il},
  \end{aligned}
\end{equation}
where $l$ labels the retained Schmidt coefficients and remains a \emph{classical} index carried in $S_{ll}=s_l$.
To continue evolution in the quantum-index layout, this classical index is converted to a quantum index after each truncation.
For a hybrid site $[\mathcal{A}]_i=|\psi_i\rangle$, the required state is
\begin{equation}\label{eq:c2q}
  |\Psi\rangle=\sum_i |i\rangle|\psi_i\rangle.
\end{equation}
Figure~\ref{fig:c2q} implements the conversion for a single qubit.
The multi-qubit case is analogous.
Each $U_i$ prepares $|\psi_i\rangle$, and the circuit returns $|\Psi\rangle$.

\subsection{Observables}

With truncated sites $\widetilde{\mathcal{A}}$, $\widetilde{\mathcal{B}}$ and diagonal connector $S$, the expectation value takes the same form as Eq.~\eqref{eq:global_exp_two_cite} of the main text, with $\mathbf{c}^{-}$ replaced by the connection registers $\mathbf{q}^{-}$.
Here $q_{\rm phys}^{A}$ and $q_{\rm phys}^{B}$ label the physical qubits of sites $A$ and $B$, as distinct from their connection registers.
\begin{equation}\label{eq:global_exp_quantum}
  \begin{aligned}
     &[\langle\Psi_{AB}|\hat{O}|\Psi_{AB}\rangle]_{\mathbf{q}^{-}_A, \mathbf{q}'^{-}_A, \mathbf{q}^{-}_B, \mathbf{q}'^{-}_B} \\
    =& \sum_{l,l',q_{\rm phys}^{A},{q'}_{\rm phys}^{A},q_{\rm phys}^{B},{q'}_{\rm phys}^{B}} S_{ll}S_{l'l'}[\widetilde{\mathcal{A}}^*]^{\mathbf{q}^{-}_A,\, q_{\rm phys}^{A},\, l}[\hat{O}_A]^{q_{\rm phys}^{A}, {q'}_{\rm phys}^{A}}[\widetilde{\mathcal{A}}]^{\mathbf{q}'^{-}_A,\, {q'}_{\rm phys}^{A},\, l'}\\
    &\times[\widetilde{\mathcal{B}}^*]^{\mathbf{q}^{-}_B,\, q_{\rm phys}^{B},\, l}[\hat{O}_B]^{q_{\rm phys}^{B}, {q'}_{\rm phys}^{B}}[\widetilde{\mathcal{B}}]^{\mathbf{q}'^{-}_B,\, {q'}_{\rm phys}^{B},\, l'} \\
    =&\sum_{l,l'} s_l s_{l'}\langle\psi_{\mathbf{q}^{-}_A,l}|\hat{O}_A|\psi_{\mathbf{q}'^{-}_A,l'}\rangle
    \langle\phi_{\mathbf{q}^{-}_B,l}|\hat{O}_B|\phi_{\mathbf{q}'^{-}_B,l'}\rangle \\
    =&\sum_{l,l'} p_l p_{l'}\,\mathcal{Z}\,\mathcal{Z}\,
    \langle\psi_{\mathbf{q}^{-}_A,l}|\hat{O}_A|\psi_{\mathbf{q}'^{-}_A,l'}\rangle
    \langle\phi_{\mathbf{q}^{-}_B,l}|\hat{O}_B|\phi_{\mathbf{q}'^{-}_B,l'}\rangle,
  \end{aligned}
\end{equation}
with $\mathcal{Z}=\sum_l s_l$ and $p_l=s_l/\mathcal{Z}$.
Each shot samples $l$ and $l'$ independently.
Local overlaps on $q_{\rm phys}^{A}$ and $q_{\rm phys}^{B}$ are measured after projecting the connection registers onto $|l\rangle$ and $|l'\rangle$.
The shot accumulates the reweighted outcome $\mathcal{Z}\mathcal{Z}\,\widehat{o}_A\widehat{o}_B$.
When the free indices coincide (or for a fully contracted Hermitian expectation), Hermiticity of $\hat{O}_A$ and $\hat{O}_B$ makes each summand real, so one may accumulate $\mathcal{Z}\mathcal{Z}\,\mathrm{Re}(\widehat{o}_A\widehat{o}_B)$.

\subsection{Worst-case qubit counts}\label{sec:qubit_resources}

We compare only how many qubits are needed to hold the connector labels under the two encodings.
Take $N$ sites of uniform degree $k$ after truncation to bond dimension $\chi$, with $n_{\rm phys}$ physical qubits per site.
As above, $\boldsymbol{\lambda}$ and the truncated Schmidt coefficients in $S$ remain classical in both counts.

In the classical-index layout each hybrid site carries a multi-index of size $\chi^{k}$.
Holding every branch as a separate $n_{\rm phys}$-qubit register at once therefore costs
\begin{equation}\label{eq:qubit_classical}
  N_{\rm qubit}^{\rm (classical)} \sim N\, n_{\rm phys}\, \chi^{k}
\end{equation}
qubits in the worst case.
If the same physical registers are reused while an outer loop iterates over the classical indices, only the peak circuit width in that loop must then be available at one time.
Examples are $2n_{\rm phys}+1$ for a two-register Hadamard test of a Gram entry, and $n_{\rm phys}+\lceil\log_2 d\rceil$ for absorption of a classical tensor with a Case~2 contraction, where $d$ is the untruncated bond dimension.

In the quantum-index layout each site keeps one physical register together with $k$ connection registers of width $n_\chi=\lceil\log_2\chi\rceil$, hence
\begin{equation}\label{eq:qubit_qconn}
  N_{\rm qubit}^{\rm (quantum)} = N\bigl(n_{\rm phys}+k\lceil\log_2\chi\rceil\bigr).
\end{equation}
Relative to Eq.~\eqref{eq:qubit_classical}, the quantum encoding replaces the factor $\chi^{k}$ by an additive $k\lceil\log_2\chi\rceil$ on each site.
Sampling costs are separate and are analyzed in Sec.~\ref{sec:sampling_overhead}.

\section{Sampling overhead on multiple cuts}\label{sec:sampling_overhead}

As in the main text, a shot estimator for one free-index entry on a truncated cut takes the form $X=\mathcal{Z}^{2}\,\widehat{o}_A\widehat{o}_B$ with $\mathcal{Z}=\sum_l s_l$.
For local Pauli observables one has $|\widehat{o}_A\widehat{o}_B|\le 1$, and therefore $\mathrm{Var}(X)\le\mathbb{E}[|X|^{2}]\le\mathcal{Z}^{4}$, so $O(\mathcal{Z}^{4}/\varepsilon^{2})$ shots suffice.
After truncation and renormalization to $\sum_l s_l^{2}=1$, Cauchy--Schwarz gives $\mathcal{Z}\le\sqrt{\chi}$ and thus $O(\chi^{2}/\varepsilon^{2})$ per entry.
Both bra and ket reweighting factors multiply inside the same estimator.
For $B>1$ cuts that bound extends to a joint draw over every connector.

Suppose the network expectation is a multiple sum over connector indices on cuts $b=1,\ldots,B$.
One shot draws every required index and accumulates the product of all reweighting factors into a single estimator of the final scalar.
The shot reweighting factor is
\begin{equation}
  \mathcal{Z}_{\mathrm{shot}}=\prod_{b=1}^{B}\mathcal{Z}_b^{2},
\end{equation}
with $\mathcal{Z}_b=\sum_l s_l^{(b)}$.
For local Pauli observables each measurement product is bounded by $1$ in absolute value, so $\mathrm{Var}(X)\le\mathbb{E}[|X|^{2}]\le\mathcal{Z}_{\mathrm{shot}}^{2}$, and
\begin{equation}\label{eq:N_joint}
  N_{\rm joint}=O\!\left(\frac{\mathcal{Z}_{\mathrm{shot}}^{2}}{\varepsilon^{2}}\right)
  =O\!\left(\frac{1}{\varepsilon^{2}}\prod_{b=1}^{B}\mathcal{Z}_b^{4}\right).
\end{equation}
After truncation this is $O(\chi^{2B}/\varepsilon^{2})$.
This bound applies to readout of a network already truncated to $\chi$ on each cut.
The shot cost is exponential in $B$, but polynomial in $\chi$ once $B$ is fixed.

For quantum simulation with few hybrid sites, $B$ stays small, so joint sampling scales as a polynomial in $\chi$.
Circuit knitting has no analogous truncated-$\chi$ scaling.
Its overhead is a product over cuts.
For a remote Pauli rotation $e^{-i\theta_{ij}\sigma_i\otimes \sigma_j}$ the single-cut factor is $(1+2|\sin\theta_{ij}|)^{2}$.
The product remains exponential in the cut count for any number of sites, even when each factor is modest.
Without truncation, THTN readout recovers the same exponential cut scaling, as noted in the main text.

The scalings in this section are sampling costs for readout.
They are not an end-to-end complexity for the full distributed simulation.
Each truncation is dominated by the Gram-matrix estimate and by the Case~2 contractions that absorb the mapping tensors.
The Gram matrix has size set by the cut dimension, and its estimation cost is polynomial in $\chi$.
Once $\chi$ is fixed, the number of Case~2 (or Case~3 for quantum connecting indices) contractions and the qubit width of each circuit remain bounded.

\end{document}